\documentclass[12pt, openany, oneside]{memoir}

\usepackage[a4paper,left=2cm,right=2cm,top=2.5cm,bottom=4cm]{geometry}

\usepackage[utf8]{inputenx}
\usepackage[T1]{fontenc}
\usepackage{amsmath}
\usepackage{scalerel,amssymb}
\usepackage{needspace}

\usepackage{lmodern}           
\usepackage[scaled]{beramono}  
\usepackage{microtype}  

\usepackage{paralist}
\usepackage{booktabs}
\usepackage{hyphenat}
\usepackage{xspace}
\usepackage[all]{nowidow}
\usepackage{csquotes}

\usepackage{url}

\usepackage{graphicx}
\usepackage{imakeidx}
\usepackage[hidelinks]{hyperref}
\makeindex[intoc]

\usepackage[explicit]{titlesec}
\usepackage{tikz}

\usepackage{todonotes}
\usepackage{algorithm}
\usepackage{algorithmic}

\usepackage{etoolbox}
\newcommand{\indexauthors}[1]{%
	\forcsvlist{\index}{#1}
}

\makeoddfoot{plain}{}{}{\thepage} 
\makepagestyle{pter}
\makeevenhead{pter}{}{}{\itshape\leftmark}
\makeoddfoot{pter}{}{}{\thepage}
\makeevenfoot{pter}{\thepage}{}{}
\chapterstyle{komalike}

\makeatletter
\renewcommand{\maketitle}{\bgroup\setlength{\parindent}{0pt}
	\begin{flushleft}
		\large\bfseries{\@title}
	\end{flushleft}\egroup
}
\makeatother

\newcommand\chapterbar{%
	\tikz[baseline, trim right=1cm] {
		\node [
		fill=yellow!90,
		text = black,
		anchor= base east,
		minimum height=3.5ex] (a) at (1cm,0) {
			\textbf{\thechapter}
		};   
	}%
}
\titleformat{\chapter}{\Large\bfseries\sffamily}{\chapterbar}{0.25cm}{#1} 
\titlespacing*{\chapter}{0cm}{3.5ex plus 1ex minus .2ex}{2.3ex plus .2ex}
\titlespacing*{name=\chapter,numberless}{-0cm}{3.5ex plus 1ex minus .2ex}{2.3ex plus .2ex}

\newcommand\sectionbar{%
	\tikz[baseline,trim right=1cm] {
		\node [
		fill=yellow!90,
		text = black,
		anchor= base east,
		minimum height=3.5ex] (a) at (1cm,0) {
			\textbf{\thesection}
		};   
	}%
}

\newcommand{\sectionFormat}{\large\bfseries\sffamily}
\titleformat{\section}{\sectionFormat}{\sectionbar}{0.25cm}{{#1}}

\titleformat{\subsection}{\normalsize\bfseries\sffamily}{\sectionbar}{0.25cm}{\textcolor{black}{#1}}

\makeatletter
\renewcommand\@makefntext[1]{\leftskip=0em\hskip0em\@makefnmark#1}
\makeatother

\makeatletter
\renewenvironment{thebibliography}[1]
{\subsection*{\bibname}
	\@mkboth{\MakeUppercase\bibname}{\MakeUppercase\bibname}%
	\list{\@biblabel{\@arabic\c@enumiv}}%
	{\settowidth\labelwidth{\@biblabel{#1}}%
		\leftmargin\labelwidth
		\advance\leftmargin\labelsep
		\@openbib@code
		\usecounter{enumiv}%
		\let\p@enumiv\@empty
		\renewcommand\theenumiv{\@arabic\c@enumiv}}%
	\sloppy
	\clubpenalty4000
	\@clubpenalty \clubpenalty
	\widowpenalty4000%
	\sfcode`\.\@m}
{\def\@noitemerr
	{\@latex@warning{Empty `thebibliography' environment}}%
	\endlist}
\makeatother

\newcommand\eg{e.\,g.\xspace}

\newcommand\wrt{w.\,r.\,t.\xspace}

\newtheorem{characterization}{Characterization}
\newcommand{\lvar}[1]{\text{#1}}

\newenvironment{conf-abstract}[4][]{
	{\renewcommand\textsuperscript[1]{}
	 \renewcommand{\\}{}
		\noindent
		\section[#2 \textit{(#3)}]{}
		\label{#1}
	}
		\begin{center}
			{\sectionFormat #2\par}
			\normalfont {#3}\par
			#4
		\end{center}
	
	\setcounter{footnote}{0}
	\bigskip
}{
	\bigskip
}

\begin{document}

\frontmatter
\thispagestyle{empty}

\vspace*{10em}

{
	\fontsize{24}{32}
	\bfseries 
	\boldmath
	\sffamily
	\noindent 
	Explainable Software for \\ Cyber-Physical Systems (ES4CPS) \par
}

\vspace{4em}

{
	\fontsize{16}{19}
	\bfseries 
	\boldmath
	\sffamily 
	\noindent 
	Report from the GI Dagstuhl Seminar 19023\\January 06--11 2019, Schloss Dagstuhl\par
}

\vspace{8em}

{
	\normalsize
	\boldmath
	\sffamily 
	\noindent 
	Edited By:
	\par
}

{
	\fontsize{16}{19}
	\bfseries 
	\boldmath
	\sffamily 
	\noindent 
	Joel Greenyer \\
	Malte Lochau\\
	Thomas Vogel
	\par
}

\thispagestyle{empty}

\vspace*{13em}

\sffamily
\noindent 
\emph{Editors} \\[0.2cm]
\hspace*{-3mm}
\begin{tabular}{lll}
	Joel Greenyer \\
	Software Engineering for Ubiquitous Applications\\
	Leibniz Universität Hannover, DE\\	
	\texttt{greenyer@inf.uni-hannover.de} \\
	\\	
	Malte Lochau \\
	Real-time Systems Lab\\
	TU Darmstadt, DE\\
	\texttt{malte.lochau@es.tu-darmstadt.de}\\
	\\
	Thomas Vogel   \\
	Software Engineering Research Group \\ 
	Humboldt-Universität zu Berlin, DE\\ 
	\texttt{thomas.vogel@informatik.hu-berlin.de}\\
\end{tabular}

\bigskip
\bigskip
\bigskip
\bigskip

\noindent 
\emph{Published online with open access}\newline
\bigskip
\emph{Publication date}
(April, 2019)

\bigskip

\noindent 
\emph{License}\\[0.2cm]
\includegraphics[width=0.75\marginparwidth]{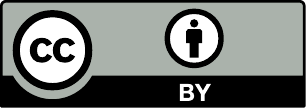}\newline
This work is licensed under a Creative Commons Attribution 4.0 International (CC~BY~4.0):\\ \url{https://creativecommons.org/licenses/by/4.0/legalcode}.\\
In brief, this license authorizes each and everybody to share (to copy, distribute and transmit) the work under the following conditions, without impairing or restricting the authors' moral rights:
\vspace{-1em}
\begin{itemize}
	\item Attribution: The work must be attributed to its authors.
\end{itemize}

\title{Report from the GI Dagstuhl Seminar 19023 \\
Explainable Software for Cyber-Physical Systems (ES4CPS)}

\maketitle

\bigskip

\noindent
\textbf{Edited by}\\
Joel Greenyer\textsuperscript{1}, 
Malte Lochau\textsuperscript{2}, 
and Thomas Vogel\textsuperscript{3}
\ \\ 
\ \\\textsuperscript{1} Leibniz Universität Hannover, DE, {greenyer@inf.uni-hannover.de} \vspace*{1mm}
\ \\\textsuperscript{2} TU Darmstadt, DE, {malte.lochau@es.tu-darmstadt.de}\vspace*{1mm}
\ \\\textsuperscript{3} Humboldt-Universität zu Berlin, DE, thomas.vogel@informatik.hu-berlin.de

\bigskip

\noindent
\textbf{Abstract}\\
\normalfont
This report documents the program and results of the GI-Dagstuhl Seminar 19023 on Explainable Software for Cyber-Physical Systems (ES4CPS)\footnote{\url{https://thomas-vogel.github.io/ES4CPS/}}. The seminar was concerned with the challenge that for future Cyber-Physical Systems (CPS), it will become increasingly relevant to explain their behavior (past, current, and future behavior, why a certain action was taken, how a certain goal can be achieved, etc.) to users, engineers, and other stakeholders. In order to increase the explainability of CPS and their engineering tools, fundamental, interdisciplinary research is required; solutions from multiple disciplines within software engineering, systems engineering, and related fields have to be applied, combined, and researched further. The goal of this seminar was to serve as a starting point for an interdisciplinary coordination of research activities targeting ES4CPS and an incubator of a new research community around this topic.

\bigskip

\noindent
\textbf{2012 ACM Subject Classification}: D.2.10 [Software Engineering] Design, D.2.11 [Software Engineering] Software Architectures


\newpage
\tableofcontents*

\mainmatter

\chapter[Executive Summary \normalfont\textit{{(Joel Greenyer, Malte Lochau, Thomas Vogel)}}]{Executive Summary}

{Joel Greenyer\textsuperscript{1}, 
	Malte Lochau\textsuperscript{2}, 
	and Thomas Vogel\textsuperscript{3}}
\indexauthors{Greenyer!Joel, Lochau!Malte, Vogel!Thomas}

\ \\\textsuperscript{1} Leibniz Universität Hannover, DE, {greenyer@inf.uni-hannover.de} \vspace*{1mm}
\ \\\textsuperscript{2} TU Darmstadt, DE, {malte.lochau@es.tu-darmstadt.de}\vspace*{1mm}
\ \\\textsuperscript{3} Humboldt-Universität zu Berlin, DE, thomas.vogel@informatik.hu-berlin.de

\bigskip
\noindent

\section{Motivation}

Collaborating and autonomously driving cars, smart grids, as well as modern systems in industry (Industry 4.0) or health care are examples of communicating embedded systems where software enables increasingly advanced functionality. These novel kinds of (software) systems are frequently summarized under the term \textit{cyber-physical systems} (CPS)~\cite{summary:Broy2010,summary:Lee2008}. CPS are usually described by three essential characteristics: CPS perform complex computations, CPS conduct control tasks involving discrete and continuous data and signal-processing, and CPS are (parts of) distributed, and even mobile, communication systems.

As a consequence, CPS become more and more complex for several reasons. Among others, because (1) they control increasingly complex processes in the physical world, (2) due to the distributed, concurrent, and hybrid nature of their software, (3) because of changing system topologies (e.g., moving cars, reconfigurable smart factory components, suppliers and consumers join or disconnect from electrical grids), and, last, (4) because the systems often learn and continuously adapt themselves to ever-changing contexts and environmental conditions.

This increasing complexity poses several challenges throughout all software development and analysis phases, but also during their usage and maintenance. In particular, it becomes increasingly difficult for system and software engineers, but also users, insurers, lawyers, auditors, etc., to comprehend the behavior of a system, especially in the case of software that relies more and more on learning and self-adaptive functionality. Why did the system respond in a certain way? How will the system react to certain inputs? How can the system achieve a certain goal? What are the reasons for an observed failure of the system and how can this behavior be reproduced?---Being able to answer these questions is important, especially for safety-critical systems, so that, (1) during development, engineers can effectively ensure the quality of the system, and (2) during operation, users can develop trust towards reliability of their systems. 

It will thus be increasingly relevant for future CPS to explain their behavior (past, current, and future behavior, why a certain action was taken, how a certain goal can be achieved) to users and other stakeholders, like lawyers and insurers in a graspable and dependable way. To this end, it will be pivotal for the different application domains of CPS and their respective engineering tools and techniques to be able to infer, update, document, and provide such explanations during different stages of system development and the system’s operation. We use the term \textit{explainability}. to describe the capability of both the system and its engineering tools to explain certain aspects of interest about the system, both in a human-comprehensible and machine-processable format. In order to increase the explainability of current and future CPS and their engineering tools, fundamental, interdisciplinary research is required; solutions from multiple disciplines within software engineering, systems engineering, and related fields may have to be applied and combined, for example:

\begin{itemize}\itemsep0em
	\item Model-based Development
	\item Requirements Engineering
	\item Formal Methods
	\item Dependable Systems Engineering
	\item Control Engineering
	\item Testing
	\item Simulation
	\item Visualization
	\item Product-Line Engineering
	\item Software Architecture
	\item Self-Adaptive Systems
	\item Human-Computer Interaction
	\item Machine Learning
	\item Organic Computing
	\item Areas of application: Robotics, Automotive Systems, Health Care Systems, etc.
\end{itemize}

We observe that research related to Explainable Software for Cyber-Physical Systems (ES4CPS) is indeed conducted in the different communities, but the research is currently only weakly linked and there are no venues where an appropriate interdisciplinary coordination of research activities focused on ES4CPS takes place.

\section{Seminar Goals}

The goal of this GI-Dagstuhl Seminar was to serve as a starting point for an interdisciplinary coordination of research activities targeting ES4CPS. The seminar should serve as an incubator of a new research community around this topic. Specific sub-goals were: (1) Build a network of young researchers working on the topic of ES4CPS; (2) Facilitate cross-disciplinary insights into research activities; (3) Support researchers in identifying common goals, intersections, and connection points in their research; (4) Foster cooperation of researchers leading to writing joint papers or project proposals.

\section{Program and Results}

The 5-day seminar was structured to have position talks of the participants as well as keynotes by invited speakers for two and a half days. Keynotes were given by Mohammad Mousavi on \textit{Conformance Testing as a Tool for Designing Connected Vehicle Functions}, by Holger Giese on \textit{Challenges for Engineering Smart Cyber-Physical Systems}, and by Holger Hermanns on \textit{Foundations of Perspicuous Software Systems}. For the rest of the seminar, we identified special interest and formed working groups to refine these specific interests. We formed the following four working groups:

\begin{itemize}\itemsep0em
	\item On Explainability and its Characterization 
	\item Self-Explainable Cyber-Physical Systems
	\item Explainability and Testing
	\item Explainability for Safety and Security
\end{itemize}


\section*{Acknowledgments}
We would like to thank the Gesellschaft for Informatik e.V. (GI, German Society for Informatics) and Schloss Dagstuhl for partially funding this seminar.

\chapter{Breakout Groups}
\begin{conf-abstract}[]
	{On Explainability and its Characterization}
	{Dimitri Bohlender\textsuperscript{1},
	    Francisco J. Chiyah Garcia\textsuperscript{2},
		Maximilian Köhl\textsuperscript{3},\\ 
		Claudio Menghi\textsuperscript{4}, and
		Andreas Wortmann\textsuperscript{5}}
	{\textsuperscript{1} RWTH Aachen University, Aachen, Germany, {bohlender@embedded.rwth-aachen.de}\\
	\textsuperscript{2} Heriot-Watt University, Edinburgh, UK, fjc3@hw.ac.uk\\
	\textsuperscript{3} Saarland University, Saarbrücken, Germany, mkoehl@cs.uni-saarland.de\\
	\textsuperscript{4} University of Luxembourg, Luxembourg, Luxembourg, claudio.menghi@uni.lu\\
	\textsuperscript{5} RWTH Aachen University, Aachen, Germany, wortmann@se-rwth.de}
	\indexauthors{Bohlender!Dimitri, Garcia!Francisco J. Chiyah, Köhl!Maximilian, Menghi!Claudio, Wortmann!Andreas}
\noindent
The goal of this breakout-group was to provide a characterization of \emph{explainable systems} that establishes a common terminology to be used within the field, enables classification of existing and future work, and explicates the connections underlying different strands of work.
Across the seminar we noticed different usages of the term \emph{explainability} coming from different researchers and research groups.
This lack of a common ground complicates exchange and comparison of research, generates confusion, and adds unnecessary communication overhead.
To overcome those problems, we provide a unifying characterization we envision to be adopted by different researchers and research groups.

Our goal was not to provide a technical definition of \emph{explainable systems} but a general conceptual framework.
For concrete application domains, this framework needs to be instantiated by technical definitions.
See Section~\ref{sec:testing_samples} for examples on how to use the characterization and \cite{g1:bohlender2019characterization} for a more details on the proposed characterization.


Besides building upon existing characterizations and definitions presented during the workshop, we reviewed additional literature \cite{g1:sep-scientific-explanation,g1:sep-reasons-just-vs-expl} and consulted various online resources, such as dictionaries (\eg,~\cite{g1:Oxford:explanation}).
Throughout this process, we noticed that there was no characterization that was general enough to accommodate all the work presented during the seminar and fits our purposes.
Thus, we developed such a characterization in an iterative processes by discussing it with other participants.

In what follows, we introduce and discuss this characterization on the basis of the following running example: we consider an automotive navigation system entrusted to plan the temporally shortest route to the driver’s home.
However, at some point, the computed route leads off the freeway just to reenter it again at the next possibility, effectively bypassing a section of the freeway on a more speed-limited road.
This route may seem implausible to the driver and begs the question why the navigation system does not favor staying on the freeway.
In the context of CPS, the need for explanations originates in such lacks of understanding.
A lack of understanding may concern some event happened within a system, a decision of a system, the behavior of a system under certain circumstances, the question how to achieve a specific goal by using a system, or some other aspect, quality, or property of a system.
We call what is to be explained the \emph{explanandum} \cite{g1:sep-scientific-explanation}.


Before presenting our characterization, we introduce its underlying core concepts.
\paragraph{Target group and representative agent.}
To make sense of an alleged explanation one needs certain background knowledge and other cognitive capacities.
In the context of our example, imagine that the reason for the seemingly implausible routing is the system’s reaction to a message about an accident on the freeway, received via the Traffic Message Channel (TMC).
While the TMC message log would be a perfect explanation for an engineer, it is too low-level and verbose to be an explanation for the average human driver.
Hence, what counts as an explanation is relative to a specific agent or a group of agents.
To generalize and abstract away from concrete agents we employ the notion of a \emph{target group} like \emph{human drivers} or \emph{automotive engineers}.
Note, this introduces vagueness but enables generalization.
Still, a target group may contain an agent who lacks the required abilities and knowledge.
Referring to our running example, there may be a driver who is inexperienced with reading maps or does not understand a certain symbol indicating an accident.
To avoid such corner cases our characterization refers to \emph{representative agents}.
Representative agents are assumed to be equipped with the relevant background knowledge and cognitive capacities.

\paragraph{Information and representation.}
Intuitively, an adequate explanation carries the information necessary for understanding.
In our example, the information necessary for understanding the seemingly implausible route is that there is an accident on the bypassed freeway segment.
This information can be represented in different ways, for instance, using a symbol on the map or by displaying a text message.
Importantly, the complete TMC message log also carries this information.
Clearly, presenting the \emph{same} information in different forms leads to explanations with different qualities.
Depending on the target group a different representation of the same information may be applicable.
Separating information and representation enables the comparison of approaches which produce different representations carrying the \emph{same} information.
The process of producing an explanation comprises two phases.
First the relevant information needs to be gathered, \eg, by monitoring the system, and, second, this information needs to be transferred into a suitable representation.


Given these core concepts we characterize \emph{explanation} as follows.


\begin{characterization}[Explanation]
	\label{def:explanation}
	A representation $E$ of some information $I$ is an \emph{explanation} of explanandum $X$ with respect to target group $G$ iff
	the processing of $E$ by any representative agent $A$ of $G$ makes $A$ understand $X$.
\end{characterization}

%

Intuitively, what makes a system explainable \wrt to a target group is access of the group to explanations.
To provide access to explanations they need to be produced by something or someone.
This \emph{means for producing} an explanation could be another system, the system itself, or even a human expert.
The mere theoretical existence of some explanation is, however, not sufficient for a system to be explainable.
We lift the characterization of \emph{explanation} in order to specify what it takes for a system to be \emph{explainable}:

\begin{characterization}[Explainable System]
	A system $S$ is \emph{explainable} by means $M$ with respect to some explanandum\footnote{Here $X$ is not an arbitrary explanandum but an explanandum related to $S$.} $X$ and target group $G$ iff $M$ is able to produces an $E$ such that $E$ is an explanation of $X$ \wrt $G$.
\end{characterization}

We evaluated our characterization through discussions with other breakout-groups.
These discussions were conducted with the purpose of refining the characterization and eliminating potential flaws.
The discussions also revealed the benefits of using the final characterization we presented.
In the following, we recapitulate the most important insights.


\emph{Relying on the notion of target groups allows an effective classification of works depending on the intended recipients of an explanation.} 
As we have seen throughout the workshop, some approaches are targeted towards average end-users while others require more specific background knowledge and cognitive capacities.
Further, identifying the intended recipients for explanations and ensuring that those actually understand the relevant explananda after processing an explanation is an essential a part of engineering explainable systems.
Our characterization makes this need explicit by demanding from a domain-specific instantiation that it clarifies to which target group what needs to be explained.


\emph{Separating information and representation allows discriminating between the explanation and the information needed to produce it.} 
Drawing a sharp distinction between representation and information allows the classification of research according to four research questions: (i) Which information is necessary for understanding? (ii) How to extract the necessary information, \eg, by monitoring? (iii) Which representation to choose for which target group, and (iv) how to produce the required representation based on the extracted information?
\emph{Characterizing the notion of explanation using the notions of processing and understanding is indeed illuminating.}
One may argue that this characterization merely is a shift in terminology which is not very illuminating with regard to what an explanation is.
This objection demands to further clarify what processing and understanding amounts to.
Depending on the target group, the authority of what it means to understand falls into different hands, and so does the authority of what counts as an explanation.
Part of designing an explainable system is to make sure that alleged explanations make the relevant target group understand the explanandum, that is, that they are indeed explanations of the explanandum according to our characterization, and not something else.
To this end, it is advisable to turn to psychology and the cognitive sciences for further interdisciplinary research.
Our characterization does not inadequately anticipate the outcome of such research but makes its necessity explicit.
It separates different concerns: what it means for a human agent to understand something is not a primary question computer scientists shall be concerned about. It is a question that has to be answered in collaboration with psychology and the cognitive sciences.
Existing research in the area of explainable systems is very broad.
Nevertheless, this research rightly claims to provide explanations and make systems explainable.
The link between those lines of work is not precise and technical but instead rooted in the vague concept of understanding.

As a conclusion, based on our evaluation, we believe that adopting the proposed characterization is beneficial to improve communication, foster collaboration, and compare different lines of works.
A central result of our investigation is that explainability is not an absolute requirement but relative. A system is not just explainable \emph{per se} but only with respect to certain aspects and to certain groups.
As we have shown, our characterization establishes a common terminology to be used within the field, enables the classification of existing and future work, and explicates the connections underlying different strands of work.


\end{conf-abstract}\newpage
\begin{conf-abstract}[group:self-explainable-systems]
{Self-Explainable Cyber-Physical Systems}
{Mathias Blumreiter\textsuperscript{1}, 
Joel Greenyer\textsuperscript{2}, 
Francisco J. Chiyah Garcia\textsuperscript{3},\\
Jesper Karlsson\textsuperscript{4},
Verena Klös\textsuperscript{5}, 
Maike Schwammberger\textsuperscript{6},\\ 
Christoph Sommer\textsuperscript{7}, 
Andreas Vogelsang\textsuperscript{8}, and 
Andreas Wortmann\textsuperscript{9}}
{\textsuperscript{1} Hamburg University of Technology, Hamburg, Germany, mathias.blumreiter@tuhh.de\\
\textsuperscript{2} Leibniz Universität Hannover, Germany, greenyer@inf.uni-hannover.de\\
\textsuperscript{3} Heriot-Watt University, Edinburgh, UK, fjc3@hw.ac.uk\\
\textsuperscript{4} KTH Royal Institute of Technology, Sweden, jeskarl@kth.se\\
\textsuperscript{5} TU Berlin, Berlin, Germany, verena.kloes@tu-berlin.de\\
\textsuperscript{6} University of Oldenburg, Oldenburg, Germany, schwammberger@informatik.uni-oldenburg.de\\
\textsuperscript{7} Paderborn University, Paderborn, Germany, sommer@ccs-labs.org\\
\textsuperscript{8} TU Berlin, Berlin, Germany, andreas.vogelsang@tu-berlin.de\\
\textsuperscript{9} RWTH Aachen University, Aachen, Germany, wortmann@se-rwth.de}
\indexauthors{Blumreiter!Mathias, Greenyer!Joel, Garcia!Francisco J. Chiyah, Karlsson!Jesper, Klös!Verena, Schwammberger!Maike, Sommer!Christoph, Vogelsang!Andreas, Wortmann!Andreas}

\noindent
The goal of this breakout-group was to examine the potential of a Cyber-Physical System to explain itself rather than being explained by some tool or engineer from the outside. The advantage of such a self-explainable system is that an explanation can be delivered whenever something needs to be explained, e.g. particularly at run-time.

Such a need for an explanation during run-time can be either detected by the system itself or triggered by a user (end-user, engineer, lawyer,...), or even by other cooperating systems. In general, the goal of an explanation is to explain a certain (expected/ unexpected) behaviour of the system. An example for a user-question for an autonomous braking system could e.g. be ``Why is the car braking?''.

However, generating the explanation for such a behaviour by itself is a challenging task for a system. It requires the system to have some understanding (i.e. model) of itself and its requirements, but also of its surrounding environment and the recipient of the explanation.

In the example there are several possible explanations for the behaviour ``braking''. Explanations e.g. could be
\begin{itemize}\itemsep0em
	\item ``Because the road ahead is in poor condition.''
	\item ``Because there exists a speed limit for this section of the road.''
	\item ``Because we need to change onto the next exit lane.''
\end{itemize}
To find the correct explanation, the system needs to thoroughly analyse its own behavioural model.

We observed that there exists neither a reference framework nor a design methodology for self-explaining systems. We thus gathered ideas from our divers insights in different fields of computer science, i.e. requirement engineering, model-driven engineering, formal models and self-adaptive systems, and started formulating a respective framework for building self-explainable systems.

The promising result of our break-out group is a joint paper about our reference framework that we are currently writing. In the paper, we also present an application of our methodology to an example from autonomous driving. The paper runs by the working title ``Towards Self-Explainable Cyber-Physical Systems'' 
and is about to be submitted in spring 2019.

\end{conf-abstract}\newpage
\begin{conf-abstract}[sec:testing_samples]
{Explainability and Testing}
{Christopher Gerking\textsuperscript{1},
    Ilias Gerostathopoulos\textsuperscript{2},
	Jesper Karlsson\textsuperscript{3},\\
	Narges Khakpour\textsuperscript{4},
	Malte Lochau\textsuperscript{5},
	Lars Luthmann\textsuperscript{6},\\
	Mohammad Reza Mousavi\textsuperscript{7},
	Christoph Sommer\textsuperscript{8}, and
	Thomas Vogel\textsuperscript{9}}
{\textsuperscript{1} Paderborn University, Paderborn, Germany, christopher.gerking@upb.de\\
\textsuperscript{2} Technical University Munich, Germany, gerostat@in.tum.de\\
\textsuperscript{3} KTH Royal Institute of Technology, Sweden, jeskarl@kth.se\\
\textsuperscript{4} Linnaeus University, Sweden, narges.khakpour@lnu.se\\
\textsuperscript{5} TU Darmstadt, Darmstadt, Germany, malte.lochau@es.tu-darmstadt.de\\
\textsuperscript{6} TU Darmstadt, Darmstadt, Germany, lars.luthmann@es.tu-darmstadt.de\\
\textsuperscript{7} University of Leicester, UK, mm789@leicester.ac.uk\\
\textsuperscript{8} Paderborn University, Paderborn, Germany, sommer@ccs-labs.org\\
\textsuperscript{9} Humboldt-Universität zu Berlin, Berlin, Germany, thomas.vogel@informatik.hu-berlin.de}
\indexauthors{Gerking!Christopher,
    Gerostathopoulos!Ilias,
	Karlsson!Jesper,
	Khakpour!Narges,
	Lochau!Malte,
	Luthmann!Lars,
	Mousavi!Mohammad Reza,
	Sommer!Christoph,
	Vogel!Thomas}

\noindent
In the context of quality assurance of cyber-physical software systems in general,
and in the context of software testing in particular, we observe a particularly strong need for explainability
in many different ways.
In particular, a frequently reoccurring challenge in software testing is to conclude from a software \emph{fault}
detected by a failed test-case execution on an implementation under test
to the respective \emph{failure} within the software actually causing that fault, and
to further trace back to the potential \emph{error} which originally introduced
that failure.

In this regard, the discussions in the break-out group on explainability
and testing were centered around three main aspects, namely:
\begin{enumerate}\itemsep0em
	\item to explore the different dimensions of---and relationships between---testing
	disciplines on the one hand, and the different aspects and challenges of explainability on the other hand,
	\item to characterize sample testing scenarios with respect to explainability, and
	\item to sketch major open challenges and future research
	in the field of testing with respect to explainability.
\end{enumerate}
In the following, we will briefly summarize the key insights and conclusions gained from the
discussions on these three points, which may serve as a road-map for future research.

\subsection*{Dimensions}
We first clustered the different aspects of testing and explainability into five orthogonal
dimensions: (1) design for explainability, (2) explaining expected behaviors, (3) explaining faults,
(4) explaining applicability of testing methods, and (5) explaining non-functional behaviors.
In the following, we will briefly summarize the different topics related to these dimensions.

\paragraph*{Design for explainability.}
Concerning design for explainability, we observed several
correspondences to the requirements requested by
respective quality criteria on software design on \emph{testability} as well as
the SERP taxonomy~\cite{g3:Petersen2014finding}.
Those requirements include:
\begin{itemize}\itemsep0em
	\item controllability and observability of the implementation under test,
	\item isolateability and separation of concerns of a particular component under test,
	\item understandability of a specifiation and/or an implementation under test in terms of documentation and or self-explanation capabilities,
	\item automatability of testing, and
	\item heterogeneity concerning the imposed diversity of testing methods and tools.
\end{itemize}
Here, understandability as well as comprehensibility may
be seen as prerequisites for explainability.

In the context of embedded and cyber-physical systems, model-based
testing practices are of particular relevance.
These black-box testing approaches are even more challenging than white-box testing
in the context of explainability as the implementation source code
is often either not accessible are even not available.
In addition, the key role of inherently uncertain environmental behaviors of cyber-physical
systems usually require complex X-in-the-loop testing scenarios.
In addition, effects resulting from the inherent distribution of those
systems such as asynchronous and lossy communication
channels must be further taken into account.

All these challenges on the design for testability also impact
diagnosis and interface design of those systems in order
to foster controllability, observability and beyond.

\paragraph{Explaining expected behaviors.}
A comprehensible specification of the expected behaviors
of an implementation under test constitutes a further dimension of
the explainability challenge in testing.
Amongst others, we identified the following research areas being
of particular relevance.
First, the oracle problem is closely related to the problem of
evaluating and explaining expected behaviors of test executions.
Similarly, proper traceability information between artifacts and tests
is pivotal for testing and the explainability of expected test results.
In the context of testing cyber-physical systems, current
research trends comprise simulation-based testing and gamification of testing
as well as testing in the context of natural language processing.
Further topics include, for instance, the relationship between test results and
notions of trust and concepts for dealing with false positives and flakiness of test results.
Finally, explanations of expected behaviors become even more important
in case of regression-testing scenarios of continuously evolving systems.

\paragraph{Explaining faults.}
Analogously to explaining behaviors expected for successful test runs, respective
capabilities for explanations of errors are required in case of failed test runs.
This includes different state-of-the-art test-result
analysis techniques like slicing, delta-debugging and fault diagnosis as well as more
advanced concepts like causality-based reasoning.
As a major challenge in this context, novel techniques are
required for reasoning about one or even all possible causes of an observed fault
by solely considering a given (failed) test-case execution, which is usually
given as a plain execution trace or input/output-value pair.
Furthermore, in case of ambiguous or non-deterministic test results, approaches for
systematic refinement of specifications can be applied which are often
based on additional information extracted from failed test-case executions.
As one step further, recent approaches devise applications of automated learning approaches
having failed test cases as inputs, for instance, in the context of automated program repair.

\paragraph{Explaining applicability of testing methods.}
A wide range of powerful testing methods and tools are available
in practice, whereas a proper recommendation of their effective usage
is often an open question.
Hence, recommendations of applicability of testing methods and tools
for a given testing challenge are a further open issue directly touching
the challenge of explainability in testing.
The various different factors influencing applicability of testing methods and tools
comprise the context, the stakeholders and development processes, the available information
about the testing objectives and requirements, and the degree of testability of the
implementation under test.

\paragraph{Explaining non-functional behaviors}
Most recent testing techniques are usually limited
to functional correctness, whereas properly expressive
specification formalisms and testing techniques for analyzing critical non-functional behaviors,
as especially apparent in cyber-physical systems, are further open issues.
In particular, the interplay between cyber- and physical aspects require
hybrid modeling formalisms and corresponding testing techniques, involving real-time behaviors, continuous
signal processing, control theory and probabilistic models as well as communication networks, but also
extra-functional aspects such as security and trust.

\subsection*{Sample Scenarios}
Based on the generic definition of explainability developed
by the respective break-out group, we considered possible
concrete instantiations of the definition in terms of
sample scenarios from different application domains of testing.
In this regard, we first refined the definition to further incorporate
a dedicated \emph{goal} that drives the tester/developer
while seeking for a particular explanation.
In this way, we are able to measure the quality of explanation by means of
the improvements achieved by the tester/developer based on the provided explanation
(e.g., removing bugs or optimizing a non-functional property).

In particular, we considered four sample scenarios as will be described
in the following.

\paragraph{Explaining applicability of testing methods.}
The goal is to explain to what extent a particular testing method
or tool allows for efficiently and effectively testing a particular system under test
(i.e., to support pretesting activity by selecting an appropriate testing tool).

\begin{itemize}\itemsep0em
	\item Information $E$: context, testability of the system under test.
	\item Representation $Y$: taxonomy for testability (SERP~\cite{g3:Petersen2014finding}) showing evidence of the selected methods and tools.
	\item Phenomenon $X$: ensure quality of a system under test by an effective and efficient testing processes.
	\item Target group $G$: testers, test managers.
\end{itemize}

\paragraph{Explaining failing tests.}
The goal is to explain why a test-case execution failed (i.e., it does not produce
the expected behavior) in white-box test setting.

\begin{itemize}\itemsep0em
	\item Information $E$: test suite, test result, test-execution trace.
	\item Representation $Y$: minimal difference between passing and failing test run (e.g., a slice containing
	the line of code where the test run failed).
	\item Phenomenon $X$: a failed functional test for a white-box program unit under test---what is the cause?
	\item Target group $G$: developers, managers.
\end{itemize}

\paragraph{Explaining black-box testing.}
The goal is to explain a failed test-case execution in a black-box test setting
(e.g., testing a GUI of an app).

\begin{itemize}\itemsep0em
	\item Information $E$: test suite, specification, observable test output.
	\item Representation $Y$: video explaining a faulty usage scenario of a GUI.
	\item Phenomenon $X$: a failed test case (causing a crash or an unspecified/unexpected output).
	\item Target group $G$: developers, managers.
\end{itemize}

\paragraph{Explaining model-based testing of configurable real-time systems.}
The goal is to explain critical best-case/worst-case execution-time behaviors
of a configurable real-time system using model-based test-case generation.

\begin{itemize}\itemsep0em
	\item Information $E$: behavioral specification of real-time behaviors and a configuration model.
	\item Representation $Y$: a timed trace and a (partial) configuration enabling
	best-case/worst-case execution-time behaviors.
	\item Phenomenon $X$: ensure real-time critical minimum/maximum delay behaviors of a configurable
	implementation under test.
	\item Target group $G$: developers, tester.
\end{itemize}

\subsection*{Future Research and Open Challenges}
To conclude, we summarized several most recent open challenges
for future research on testing and explainability.

\begin{itemize}\itemsep0em
	\item How to quantify test results to explain non-functional properties?
	\item How to explain lossy timing channels through counterexamples?
	\item How to compensate (stochastic) network communication delays when explaining test results?
	\item How to design/learn explainable environment models and fault models?
	\item How to explain non-reproducible test execution results (e.g., in case of non-determinism and concurrency)?
	\item How to explain test results in cases where a-priori
	specifications are impossible (e.g., in the context of machine learning)?
\end{itemize}

As a next step, we identified the work of Halpern and Pearl~\cite{g3:Halpern2005causes,g3:Halpern2005explanations}
on counter-factual reasoning about causality and explanations as a suitable meta theory
for explainability and testing.
Based on this theory, we plan to develop a formal framework for
explaining test execution results for cyber-physical systems (e.g., involving real-time behaviors).

\end{conf-abstract}
\newpage

\begin{conf-abstract}[]
{Explainability for Safety and Security}
{Markus Borg\textsuperscript{1}, 
Simos Gerasimou\textsuperscript{2}, 
Nico Hochgeschwender\textsuperscript{3}, 
Narges Khakpour\textsuperscript{4}}
{\textsuperscript{1} RISE Research Institutes of Sweden AB, Lund, Sweden, markus.borg@ri.se\\
\textsuperscript{2} University of York, UK, simos.gerasimou@york.ac.uk\\
\textsuperscript{3} German Aerospace Center (DLR), Germany, Nico.Hochgeschwender@dlr.de\\
\textsuperscript{4} Linnaeus University, Sweden, narges.khakpour@lnu.se}
\indexauthors{Borg!Markus, Gerasimou!Simos, Hochgeschwender!Nico, Khakpour!Narges}

\noindent
Cyber-Physical Systems (CPS) are built from, and depend upon, the seamless integration of sensing, computation, control and networking components into physical objects and infrastructure, connected to the internet and to each other~\cite{g4:NIST}.
CPS used in mission- and safety-critical application domains such as automotive, manufacturing and medicine are increasingly provided with additional autonomous capabilities as a means of reducing both human input and human error.
Despite the benefits from increased autonomy, these systems should operate trustworthy and being trusted to fulfil the critical requirements associated with the domains in which they operate.

Driven by the observation that the bounds between \textit{safety} and \textit{security}, two fundamental properties for achieving trustworthy CPS, have become less clear~\cite{g4:stoneburner2006toward}, this breakout group investigated the relationship between these CPS properties and how \textit{explainability} can bridge the gap.
Safety and security are similar in the sense that each focuses on determining mitigation strategies for specific kinds of failures by analysing potential disruptions, hazards and threats.
As such, they employ similar analysis and reasoning techniques (e.g., fault trees, attack trees) for investigating failures modes, assessing their impact on the overall CPS behaviour and deriving strategies in the form of requirements or defences.

While safety and security traditionally have been considered in isolation, addressed by separate standards and regulations, and investigated by
different research communities~\cite{g4:pietre2013cross}, they are often interdependent. Frequently, sometimes focusing on one might even be at the expense of the other~\cite{g4:pietre2010modeling}.
In fact, safety and security can be at odds since safety is concerned with protecting the environment from the system whereas security is concerned with protecting the system from the environment~\cite{g4:Adelard}.
For instance, safety encourages stable software versions while security encourages frequent updates, which requires re-qualifying and re-certifying the underlying CPS. Similarly, while safety implicitly embraces easy access to CPS components for fast and easy analysis, security explicitly adopts strict authentication and authorisation mechanisms as a way to restrict access.

Recognising the similarities and differences between CPS security engineering  and safety engineering, the research community has called for increased safety and security co-engineering~\cite{g4:stoneburner2006toward, g4:sadvandi2012safety, g4:sabaliauskaite2015aligning}.
Professional organisations contribute in this effort through initiatives to join the two areas.
For instance, the International Society of Automation (ISA) established the ``Safety and Cybersecurity Division''\footnote{https://www.isa.org/division/safety/} while there are proposals to add ``security threat analysis'' to the hazard analysis mandated by the second edition of the generic safety standard IEC~61508~\cite{g4:schoitsch2016need}.

Several challenges involved in co-engineering of safety and security originate in the different roles involved. Not only do safety engineers and security engineers need to understand each other, also certification agencies must be able to comprehend the assurance cases provided by the development organization. Moreover, there are several other stakeholders, both internal and external, that should understand both design decisions related to safety and security. Examples include product managers, requirements engineers, software testers, and also to some extent the end users (drivers) and the general public.

We argue that aligning safety and security requires to overcome several engineering and communication barriers among involved stakeholders.
In particular, the cognitive gap between security and safety experts needs to be bridged to allow successful co-engineering.
We consider \textit{explainabiity} a suitable means for supporting this purpose.
In particular, we believe that explanation should become a first-class entity during a system's lifecycle, i.e. during analysis, design, implementation, execution and maintenance stages, to be able to realise explainable software that adheres to safety and security requirements.
Software explainability is a quality attribute that assesses the degree to which an observer (human or machine) can understand the reasons behind a decision taken by the system (e.g. a malicious activity, a failure or a prediction)~\cite{g4:Dam0G18}.

An explanation-aware software development process should provide different models, techniques and methodologies to model, specify, reason about and implement explanations.
To achieve this, explanations are considered as first-class entities, i.e. they are explicit in the system's lifecycle and are supported with models and techniques to be composed, abstracted away, refined, reused etc.
Explanation-related artifacts produced in different stages should conform to a conceptual framework for developing other artefacts by different stakeholders, e.g. the safety cases created by a safety engineer can be based on the same explanation models and artefacts that a forensics analyst uses to build a security threat model to perform security threat analysis.
In particular, the process should be provided with strong techniques to analyse and reason about explanations to assure, for example,  completeness, consistency and soundness.


As the first step towards this goal, the break-out group employed a meta-modeling~\cite{g4:Silva2015} approach to identify the core concepts and their relations involved in an explanation. Adopting a meta-modeling approach for developing a conceptual framework for explainability paves not only the way to harmonize all involved concepts, but also to develop tools and other means to, for example, automatically generate or validate explanations which conform to our developed meta-model. The core concepts of our meta-model are described in the following paragraphs:

\begin{itemize}
	\item \emph{Stakeholders} are the parties involved in an \emph{Explanation}. In the context of co-engineering safety and security this could be safety and security engineers. For a given explanation, each stakeholder has a certain \emph{Role}, namely either the role of an \emph{Explainer} or of an \emph{Explainee}. The former provides an explanation whereas the latter receives an explanation. It is important to note, that both \emph{Explainer} and \emph{Explainee} are not necessarily humans, but can be represented as tools or computer programs providing and creating an explanation.

	\item \emph{Stages} encode lifecycle phases and their corresponding activities. For a given explanation both the \emph{Explainer} and \emph{Explainee} are in a certain \emph{Stage}. Those are not necessarily similar, but most likely different if the safety and security process are not aligned.

	\item \emph{Explanations} are answers for questions posed by \emph{Explainees}. In order to provide those answers an \emph{Explainer} employs a \emph{Process}. Such a \emph{Process} makes use of different \emph{Artifacts} to come up with an explanation. Example artifacts are design documents, source code, standards etc. By having an explanation in place, an implicit relation between \emph{Explainer} and \emph{Explainee} is formed. This relation serves to increase -- depending on the overall \emph{Purpose} of the explanation -- trust or understandability. To ensure that an \emph{Explanation} is useful for the \emph{Explainee}, each \emph{Explanation} is associated with a \emph{Quality} attribute and corresponding \emph{Metric} to quantify this quality attribute. For example, the extent to which the \emph{Explanation} is appropriate for the \emph{Explainee}.

	\item Depending on both the \emph{Purpose} and \emph{Explainee}, an \emph{Explanation} might have different \emph{Representations}. For example, one \emph{Explainee} prefers for the sake of comprehensibility textual over graphical visualisations. It is important to note, that an \emph{Explanation} is not limited to one representation as an \emph{Explanation} might be presented to different \emph{Explainees} with different needs.
\end{itemize}

The open challenges to realise safety and security co-engineering using explanation-aware software development are:
\begin{itemize}
	\item Extending, formalizing and concretising the proposed  meta-model as the first stepping-stone to realize explanation-aware software development that facilitates co-engineering of safety and security.
	\item Identifying various quality attributes of explanations that are concerned with safety and security in addition to proposing (formal) modeling and analysis techniques to specify and analyse explanations to assure the quality attributes.
	\item Providing composition, abstraction and refinement methods to compose, abstract and refine explanations to be able to guarantee scalability and traceability of explanations that enable to link explanations (of different stakeholders) in different stages to each other.
\end{itemize}

\end{conf-abstract}

\chapter{Keynotes}
\begin{conf-abstract}[]
	{Conformance Testing as a Tool for \\ Designing Connected Vehicle Functions}
	{Mohammad Reza Mousavi}
	{University of Leicester, UK\\mm789@leicester.ac.uk}
	\indexauthors{Mousavi!Mohammad Reza}
	\noindent
Connected and Autonomous Vehicles (CAV) are taking a central position in the landscape of intelligent mobility and their rigorous verification and validation is one of the main challenges in their public deployment and social acceptance. Conformance testing is a rigorous verification technique that has been widely tried in various critical applications. In this talk, we examin the adaptations and extensions of the notions of conformance and model-based testing techniques that make them suitable for application in the CAV domain. We present how the extended techniques can be used in the design of connected vehicle functions and verify various design decisions.

\end{conf-abstract}

\begin{conf-abstract}[]
	{Challenges for Engineering Smart Cyber-Physical Systems}
	{Holger Giese}
	{Hasso Plattner Institute for Digital Engineering gGmbH\\ University of Potsdam, Germany\\{holger.giese@hpi.de}}
	\indexauthors{Giese!Holger}
\noindent
We currently can observe a transformation of our technical world into a networked technical world where besides the embedded systems with their interaction with the physical world the interconnection of these nodes in the cyber world becomes a reality. This dramatic transformation towards cyber-physical systems raises the question whether our capabilities to design, produce, and operate these future embedded systems are ready to tackle the resulting challenges. In parallel nowadays there is a strong trend to employ artificial intelligence techniques and in particular machine learning to make software behave smart. To support this trend and to unleash their full potential, cyber-physical systems must be able to operate as elements in open, dynamic, and deviating overall structures and to adapt to open and dynamic contexts while being developed, operated, evolved, and governed independently. We name the resulting cyber-physical systems to be smart as they must be self-adaptive at the level of the individual systems and self-organizing at the system of system level to cope with the emergent behavior at that level.

In this presentation we will first discuss the envisioned future scenarios for embedded systems becoming cyber-physical systems with an emphasis on the synergies networking can offer and then characterize which challenges for the design, production, and operation of these systems result. We will then discuss to what extent our current capabilities in particular concerning engineering match these challenges and where substantial improvements for the engineering are crucial. In classical engineering models are used to plan systems upfront to maximize envisioned properties on the one hand and minimize cost on the other hand. When applying the same ideas to software for smart cyber-physical systems, it soon turned out that for these systems often somehow more subtle links between the involved models and the requirements, users, and environment exist. Self-adaptation and runtime models have been advocated as concepts to covers the demands that result from these links. Lately, both trends have been brought together more thoroughly by the notion of self-aware computing systems. We will review the underlying causes, discuss some our work in this direction, and outline related open challenges and implications for future approaches to engineer such smart cyber-physical systems.
\end{conf-abstract}

 \begin{conf-abstract}[]
	{Towards a Science of Perspicuous Computing}
	{Holger Hermanns}
	{Saarland University, Saarland Informatics Campus,  Saarbrücken, Germany\\{hermanns@cs.uni-saarland.de}}
	\indexauthors{Hermanns!Holger}
	\noindent
	 From autonomous vehicles to Industry 4.0, from smart homes to smart cities – increasingly computer programs participate in actions and decisions that affect humans. However, our understanding of how these applications interact and what is the cause of a specific automated decision is lagging far behind. This problem is at the core of incidents that caught widespread media coverage, including  the diesel exhaust emissions scandal and the recent software-induced fatal  crashes of Boeing 737 MAX 8 aeroplanes. With the increase in cyber-physical technology impacting our lives, the consequences of this gradual loss in understanding are becoming severe. Systems lack support for making their behaviour plausible to their users. And even for technology experts it is nowadays virtually impossible to provide scientifically well-founded answers to questions about the exact reasons that lead to a particular decision, or about the responsibility for a malfunctioning. The root cause of the problem is that contemporary systems do not have any built-in concepts to explicate their behaviour. They calculate and propagate outcomes of computations, but are not designed to provide explanations. They are not perspicuous. The key to enable comprehension in a cyber-physical world is a science of perspicuous computing.	
	 
	 The Transregional Collaborative Research Centre 248 \href{https://perspicuous-computing.science/}{CPEC} aims at establishing the ''Foundations of Perspicuous Software Systems'' -- so as to enable comprehension in a cyber-physical world with the human in the loop. 
	 This keynote gives an overview of the scope, structure and scientific approach of \href{https://perspicuous-computing.science/}{CPEC}, which is a joint initiative of computer science researchers at \href{https://saarland-informatics-campus.de/en/}{Saarland Informatics Campus} and \href{https://tu-dresden.de/ing/informatik/}{Technische Universit\"at Dresden}, It is funded by the  Deutsche Forschungsgemeinschaft (DFG, German Research Foundation) –- project number 389792660 -- see \url{https://perspicuous-computing.science}. 
\end{conf-abstract}

\chapter{Extended Abstracts / Talks}
\begin{conf-abstract}[]
	{Refining incomplete models through system observations with NuSMV}
	{Mathias Blumreiter}
	{Institute for Software Systems\\Hamburg University of Technology, Hamburg, Germany \\mathias.blumreiter@tuhh.de}
	\indexauthors{Blumreiter!Mathias}
\noindent

\subsection*{Motivation}
Formal models have proven to be a suitable tool to tackle complexity. Hence, they are used in all engineering phases, for operation, and for documentation. However, often, the observed system behaviour  does not follow the modelled behaviour. Reasons for such discrepancies may lie in inconsistencies between high and low-level models, inconsistencies between timed and discrete models, misunderstood or vague (under-specified) requirements, or just the further development of the system while the models are not kept up-to-date. For that reason, models tend to be incomplete. Cyber-physical systems (CPS), in particular, are no exception to model incompleteness, as the complexity of their tasks provides the complexity of themselves and therewith their models. Unlikely corner-cases may be easily missed in the specification and refinement phases. Since an observation of a system's behaviour is a temporal evolution of the system's state change, temporal models may be used to capture all possible evolutions over time. The benefit of using temporal models lies in their ability to describe how the system possibly arrived in a given state as well as the ability to predict what the system may do from that point on. Hence, depending on the abstraction level (in particular w.r.t. time), different kinds of system behaviours can be assessed (e.g., safety properties, probabilistic evaluations, etc.). As the model may be incomplete, it may happen that an observation has no direct counterpart in the existing model. In such a case, the existing models are not useful any more, neither for explanation nor prediction. For that reason, we tackle the problem of automatically refining incomplete models. New observations are incrementally integrated into the model as new facts without falsifying proven knowledge (i.e., without removing former observations and originally modelled behaviour). However, the objectives of explanation and prediction are in direct conflict while integrating a new observation. In case of explanation, the observation has to be integrated such that it has the best match with known behaviours to give an idea how the observation came to be. Since such an approach introduces alternative evolutions for the matched states, the integration may also create artefacts in the model and, therewith, increase the model's imprecision. On the other hand, since the aim is to use the resulting model for prediction, the finite observation has to be integrated such that the extended model continues reasonably with known (and possibly changed) behaviour after the added observation has been evaluated. Hence, prediction requires a precise model. 

\subsection*{Refining incomplete models with NuSMV}
As a first step to refining incomplete models of cyber-physical systems, we concentrate on the extension of discrete models and discuss a realisation in the model checker NuSMV. For that, we assume that there is a model \lvar{M} given that comprises the system's behaviour known from the specification phases as well as the observations that were integrated in previous integration iterations. Furthermore, we assume a  set of properties $\Phi$ expressed in Computation Tree Logic (CTL) or Linear-Time Logic (LTL) that capture, for example, fundamental system behaviour or safety requirements. Since the system's observed behaviour may only be present in form of a partial observation, we assume a finite sequence of propositions $\tau$ that hold in the respective observed system states. An outline of the integration procedure is given in \hyperref[blumreiter:alg:algorithm]{algorithm 1}. 

\begin{algorithm}[h]
  \scriptsize
  \caption{integrate(\lvar{M}, $\Phi$, $\tau$, \lvar{Q})}\label{blumreiter:alg:algorithm}
  \begin{algorithmic}[1]
    \REQUIRE finite set of variables $\mathcal{V}$, finite model \lvar{M} over $\mathcal{V}$, set of CTL or LTL properties  $\Phi$ over $\mathcal{V}$, (finite) sequence of satisfiable propositions $\tau = \alpha_1 \rightarrow \dots \rightarrow \alpha_m$ over $\mathcal{V}$, and a quality metric \lvar{Q}
    \STATE $\mathcal{M} \gets$ matchings(\lvar{M}, $\tau$)
    \STATE $\mathcal{C} \gets \mathcal{M} \, \mathbin{\cap}$ approximate$_{\forall \phi \in \Phi}$($\phi$) 

    \FORALL {$\tau_\text{modification} \gets$ select\_best\_candidate\textsubscript{Q}($\mathcal{C}$)}
        \STATE \lvar{M\textsubscript{extended}} $\gets$ integrate(\lvar{M}, $\tau_\text{modification}$)
        \STATE (\lvar{success}, \lvar{M'}) $\gets$ repair(\lvar{M\textsubscript{extended}}, $\Phi$) \textbf{constrained by} \lvar{M} and $\tau_\text{modification}$
        \STATE \textbf{return}  ($\top$, \lvar{M'}) \textbf{on} \lvar{success}
    \ENDFOR
    
    \STATE \textbf{return} ($\bot$, \lvar{M}) 
    \ENSURE \lvar{success} $\Rightarrow$ \lvar{M'} $\vDash \tau \land \Phi$, traces(\lvar{M}) $\subseteq$ traces(\lvar{M'}), $\neg$ \lvar{success} $\Rightarrow$ \lvar{M} $=$ \lvar{M'}
  \end{algorithmic}
\end{algorithm}

We start with computing the meta-model $\mathcal{M}$ of all matchings between \lvar{M} and $\tau$. The meta-model is constructed such that it contains all traces of \lvar{M} as well as the traces that result from the possible integrations of $\tau$. Since not all of these new traces satisfy the given properties $\Phi$, we compute an overapproximation of  $\Phi$  to reduce $\mathcal{M}$ to the model of integration candidates $\mathcal{C}$. Afterwards, we iteratively select the best candidate $\tau_\text{modification}$ (according to a given quality metric \lvar{Q}) and compute the extended model \lvar{M\textsubscript{extended}}. Thereby, we integrate $\tau_\text{modification}$ such that the state space of the resulting model  is changed minimally. The reason for this minimality requirement is that we use symbolic models based on a given set of variables $\mathcal{V}$ and the integration and verification cost will increase for further integration iterations when increasing $\mathcal{V}$. Hence, we prefer to stay in a minimally changed state space with the consequence that the integration candidate may not be directly realisable in it. For that reason, we continue with an execution of our model repair algorithm. The aim is to modify  \lvar{M\textsubscript{extended}} such that the properties $\Phi$ hold while preserving the traces from the original model, the former observations, and the current observation. The repair algorithm mainly works on the artefacts created by previous iterations. In case the integration candidate is not realisable in the current state space, the repair algorithm has the option to increase the state space in a minimal way to integrate $\tau_\text{modification}$ correctly. If the integration fails even for the increased state space, the integration procedure continues with the next candidate.

\end{conf-abstract}\newpage
\begin{conf-abstract}[]
	{Explainable Restart-Behaviour of Reactive Systems Software}
	{Dimitri Bohlender}
	{Embedded Software, RWTH Aachen University, Aachen, Germany\\ {bohlender@embedded.rwth-aachen.de}}
	\indexauthors{Bohlender!Dimitri}
\noindent
\subsection*{Restart-robustness of Control Software}
Control software is at the heart of many complex systems, ranging from assembly lines to space probes, and must meet the safety and reliability requirements of its embedding system.
To this end, a lot of research has gone into formal methods for proving a program's compliance \wrt the properties of interest, such as absence of runtime errors.
However, such a proof of correctness only holds \wrt the considered specification, and only under the assumption that the program's semantics were characterised faithfully.
With controllers that are particularly tailored to certain safety-critical domains, such as programmable logic controllers (PLCs) for industrial automation, the actual hardware specifics may also affect the program's semantics in unexpected ways.

One such feature that is widespread in controllers for safety-critical applications, is non-volatile memory, or battery-backed memory areas in the case of PLCs, as this allows for controlled handling of crashes and restarts through implementation of resumption strategies.
And its not far-fetched having to account for restarts even in such applications.
For example to detect exceptional behaviour like hang, most controllers come with a \emph{watchdog timer} which restarts the hardware once it elapses, and therefore has to be reset regularly.
Furthermore, if a chemical plant experiences a power outage and switches to emergency power, to the PLCs this will look like a restart, too.

However it is typically up to the author of the control software to decide which variables to \emph{retain} in case of a restart, and make sure that the program complies with the specifications even in the context of restarts.
It turns out difficult to keep track of all the corner cases even for small examples, and justify why a choice of \emph{retain variables} is safe --
even retaining all variables is no remedy.
Accordingly, in practice, errors that only occur after program restart are a common problem in industrial control code \cite{bohlender:RestartBehaviour}.
For example consider the use case of automated drilling of holes in workpieces.
If the drill’s position or mode of operation are not retained, a restart may result in unintended movement of the drill and damage to the system itself, the payload, or persons within reach – even if such a malfunction were not possible in a restart-free operation.

To this end, we devised an approach for augmenting the characterisation of PLC program semantics with prominent implementations of writing to non-volatile memory, enabling checking a program's \emph{restart-robustness} \wrt a property of interest, given a configuration of retain variables.
To aid in the design of restart-robust control software, we also proposed a technique for synthesising configurations of retain variables that are provably safe to use \cite{bohlender:RestartRobustness}.
Since this approach is based on \emph{symbolic model checking}, we also get counterexamples, when a violation is found, or a certificate in terms of invariants, otherwise.

In terms of \emph{explainability} \cite{bohlender:bohlender2019characterization1}, we provided the \emph{means} to produce proofs, counterexamples or retain configurations (the \emph{explanations}) of the \emph{explananda} (i) whether/why a given retain configuration results in restart-robust behavior \wrt some specification and (ii) whether/how restart-robustness can be achieved in a PLC \emph{system}.
Here, developers are the \emph{target group}.

\subsection*{Challenge}
In practice, one is often interested in updating existing control software to operate ``correctly'' even in the context of restarts.
In that case, the specification is not given in some logic but as a program, and restart-robustness needs to be defined as a relational property between the nominal and the restart-augmented program behaviour.

A naïve way to define this is to require the \emph{observable} states of the restart-augmented program to be a subset of the observable states of the original program, or that this relation must be established within $k$ execution-cycles, as this would indicate whether restarting allows the program to exhibit behaviour not possible otherwise.
In contrast to this rather weak constraint, the requirement of \cite{bohlender:CrashRecoverability}, that eventually a state will be reached which is observationally equivalent to a state visited before the restart, might be too strict in practice.

So while we have the background for developing verification procedures, insights from the automation and systems engineering perspective on the feasibility of our undertaking, and reasonable definitions for relational restart-robustness, would be helpful.

\end{conf-abstract}
\newpage
\begin{conf-abstract}[]
	{Explainability First! \\Cousteauing the Depths of Neural Networks to Argue Safety}
	{Markus Borg}
	{Software and Systems Engineering Lab\\RISE Research Institutes of Sweden AB, Lund, Sweden\\{markus.borg@ri.se}}
	\indexauthors{Borg!Markus}
	\noindent
``Safety first!'' is a weary expression that has been repeated until tedium. But any organization trying to obtain a safety certification for a software-intensive system knows that safety is achieved first \textit{after} considerable effort. Hard evidence is required for safety engineers to develop a safety case, i.e., a structured argumentation for why a system is adequately safe in a specific context. Thus, in the case of Machine Learning (ML)-based Cyber-Physical Systems (CPS), a better expression would be ``Explainability first!'' –- at least in the eyes of external safety assessors.

ML-based systems will be increasingly used in safety-critical applications, e.g., in Automated Driving (AD). Increasing the level of automation has great potential to improve safety by mitigating risks introduced by humans. Human factors such as tiredness, distractions caused by smart phone usage, vivid conversations, and loud music do not affect automated vehicles. In Sweden, the most alarming studies estimate that the number of cars driven by influenced drivers  matches the number of taxis in operation\footnote{https://www.dt.se/artikel/dalarna/15-000-resor-per-dygn-med-alkoholpaverkade-forare}. While there are obvious advantages with AD, how could you tackle the challenge of developing the required safety argumentation? 

\vspace{1em}
\fbox{\begin{minipage}{0.95\textwidth}
\textbf{\#problem} We are currently studying Deep Neural Networks (DNN) trained to enable vehicle environmental perception, i.e., awareness of elements in the surrounding traffic, and how to approach the corresponding safety certification. Successful perception is a prerequisite for AD features such as lane departure detection, path/trajectory planning, vehicle tracking, and scene understanding. A wide range of sensors have been used to collect input data from the environment, but the most common approach is to rely on front-facing cameras.
\textbf{\#deeplearning \#automotive \#vision \#safetycertification}
\end{minipage}}
\vspace{1em}

DNNs have been reported to deliver superhuman classification accuracy for specific tasks, but inevitably they will occasionally fail to generalize. Unfortunately, from a safety perspective, analyzing when this might happen is currently not possible due to the black-box nature of DNNs \cite{borg:borg_safely_2019}. A state-of-the-art DNN might be composed of hundreds of millions of parameter weights, thus the approaches to verification and validation of DNN components must be different compared to approaches for human-readable source code. Techniques enforced by the current safety standard for automotive systems, ISO~26262, are not directly applicable to DNNs, e.g., source code reviews and exhaustive coverage testing.

\vspace{1em}
\fbox{\begin{minipage}{0.95\textwidth}
 \textbf{\#expertise} Apart from machine learning, we have experience in software engineering and safety-critical development. Functional safety is defined in ISO~26262 as ``absence of unreasonable risk due to hazards caused by malfunctioning behavior of electrical/electronic systems''. The fundamental assumptions are that zero risk can never be reached, but non-tolerable risks must be reduced ``as low as reasonably practicable''. The basic approach to achieve this is by developing an understanding of situations that lead to safety-related failures and, subsequently, designing the software so that such failures do not occur. Even if the system does not conform to its specification, it should not hurt anyone or anything.\\\textbf{ \#machinelearning \#requirements \#testing \#traceability}
\end{minipage}}
\vspace{1em}

Hazards, i.e., system states that might lead to an accident, play a key role in safety-critical development. In the automotive domain, hazards are typically analyzed in four steps of early requirements engineering: 1) Hazard identification to populate a hazard register, 2) Hazard assessment of probability, acceptability, and controllability, 3) Hazard analysis to identify root causes, and 4) Risk reduction by specifying safety requirements that should ensure that the system remains safe even when things go wrong. Complete traceability from safety requirements to their software (or hardware) implementation and their corresponding test cases constitute the backbone evidence of the safety case, i.e., the documented argumentation that a system is safe, provided to external safety assessors. But what happens when a trace link from a safety requirement ends in a DNN?~\cite{borg:borg_traceability_2017}

\vspace{1em}
\fbox{\begin{minipage}{0.95\textwidth}
\textbf{\#need} We need help to develop safety cases that encompass ML, i.e., complex models that are trained on massive annotated datasets. Analogous to how the French oceanographer Jacques Cousteau pioneered charting the depths of the great oceans, the software engineering community needs to dive into the state-of-the-art DNNs to explain their intricacies. We must understand how deeply we should trace from critical requirements to explain and argue safety of an ML-based system. Should we trace to source code calling an ML library? The underlying ML model? The model architecture? Its parameter weights? Or even to specific examples in the training data? We might need traceability to several of the listed examples, since safety evidence tends to aim at exhaustiveness. Surely, before any safety certification of ML-based systems can be obtained, we first need to find ways to explain why they are safe.
\textbf{ \#safetycase \#certification \#standards \#evidence}
\end{minipage}}
\vspace{1em}


\end{conf-abstract}
\newpage
\begin{conf-abstract}[]
	{Explainable Autonomy through Natural Language}
	{Francisco J. Chiyah Garcia, David A. Robb and Helen Hastie}
	{Heriot-Watt University, UK}
	\indexauthors{Garcia!Francisco J. Chiyah, Hastie!Helen}
	\noindent

Robots and autonomous systems are increasingly being deployed remotely in hazardous environments such as in the nuclear or energy sector domains \cite{garcia:Hastie2018,garcia:Jinke2017}. Typically, these remote robots instil less trust than those co-located \cite{garcia:Bainbridge2008} and thus it is important to maintain a high level of transparency regarding their behaviour. This is particularly important for robots in remote locations where they cannot be directly observed.

The obscure nature of autonomous systems makes understanding them a difficult task for non-experts. 
The lack of knowledge and transparency about how the systems operate is reflected in decreased trust and understanding, which ultimately has negative effects on the human-machine cooperation \cite{garcia:DzindoletMaryT}. Therefore, the interface between the user and the system is key to maintaining situation awareness and understanding between the system and the human operator \cite{garcia:Robb2018}.

The interfaces used to monitor autonomous systems often combine graphical elements such as the systems' updates and location so operators can understand what is happening. The complexity of these interfaces usually requires previous training and an operator overseeing the system's actions. As we move towards unmanned systems without continuous supervision, it is vital that these systems are able to communicate their current status and explain the reasoning behind their actions in a clear manner.

Explainability is an important facet of a transparent system 
as it can provide the user with a high fidelity mental model, along with increased confidence and performance \cite{garcia:LeBras18,garcia:Lim2009}.  Mental models, in cognitive theory, provide one view on how humans reason either functionally (understanding what the robot does) or structurally (understanding how it works) \cite{garcia:JohnsonLaird80}. Mental models are important as they strongly impact how and whether robots and autonomous systems are used. 
A better mental model of a system, in terms of \textit{what} the system can do and \textit{why} it is doing certain actions, can provide the user with the knowledge to understand and predict the system's behaviours. A system that is easier to predict for users is more likely to be adopted over time and a clearer mental model can also help to avoid wrong assumptions and misuse of the system.

Explanations can contribute to building accurate mental models of a system. 
For example, \cite{garcia:Lim2009} showed that explaining ``\textit{why}'' a system behaved in a certain way increased both understanding and trust, whilst ``\textit{why not}'' explanations showed only an increase in understanding, and thus both are important with regards the user's mental model. 
However, the way these explanations are given to the user can affect their benefits. As \cite{garcia:GregorBenbasat} shows, users will only take the time to process explanations if the benefits are perceived to be worth it, and do not adversely add to cognitive load \cite{garcia:Mercado16}. \cite{garcia:Kulesza2013} explored how the amount of information that the explanations give, in terms of \textit{soundness} (the level of detail) and \textit{completeness} (the amount of reasons to give), can affect the user's mental model. Thus it is important to control both what to explain and how to present it to the user.

Interaction through natural language provides an intuitive means of keeping the user informed and requires little training. Conversational agents are becoming widely used to perform tasks such as asking for the weather or restaurant recommendations. Their convenience and ability to give on-demand instant responses makes them appealing for a range of users and devices. 
Previous works have investigated natural language to provide explanations of a variety of systems including planning, deep learning models and robot rationalisation. However, autonomous systems work in a continually changing environment with dynamic goals, which cannot always be known ahead of time and preprogrammed.

The data that autonomous systems gather from the environment to decide their next action offers an insight into how they behave. Our most recent works studied how this data can be used to increase trust and understanding of an autonomous system, as well as derive explanations about its behaviour \cite{garcia:Robb2018, garcia:ChiyahHRI18, garcia:ChiyahetalINLG18}. We showed that a conversational agent running alongside the system's interface can increase the user's situational awareness and that it requires little training to understand and use. We also found that explanations with a high completeness (multiple explanations about the autonomy model) performed the best in increasing the user's understanding, whereas explanations with high soundness (high level of detail) were not as necessary. 

Future work includes how to expand the information that the Conversational Agent can process and understand, as well as deeper integration with the autonomous system. Uncertainty is a challenging issue when working with autonomous systems and therefore, understanding the best way to handle it and present it to the user is essential to increase the system's transparency. Finally, another important goal that we are working towards is the generalisation of the agent, so it can be applied to other systems regardless of the domain.

\end{conf-abstract}
\newpage
\begin{conf-abstract}[]
	{Safety Assurance in Cyber-Physical Systems: Mind the Gap}
	{Simos Gerasimou}
	{University of York, UK\\ {simos.gerasimou@york.ac.uk}}
	\indexauthors{Gerasimou!Simos}
\noindent
	Cyber-physical systems (CPSs) refer to a new generation of sophisticated
	systems that fuse the physical with the digital word~\cite{CPS}.
	CPSs are distributed, heterogeneous and interconnected systems, typically
	engineered from the seamless integration of
	computational algorithms, physical processes and communication technologies.
	CPSs have been successfully deployed in various application domains ranging
	from automotive, avionics and maritime to manufacturing, defence and medicine.
	Some illustrative examples include monitoring the safe movement of autonomous
	and non-autonomous vehicles, supporting elderly people living alone by
	detecting illnesses or accidents and raising the alarm automatically, and
	reducing energy consumption through intelligent home control
	applications~\cite{CPS}.

	While the ability of CPSs to interact with the physical word is key in
	realising effective real-word CPS applications, it also increases the
	uncertainty (both epistemic and aleatory) for engineering these systems.
	In comparison to related technological fields (e.g., embedded systems,
	mechatronics, internet of things, big data), CPSs exhibit unique
	characteristics concerning autonomy, dependability, sophisticated operation,
	and interdependencies between system components.
	These characteristics may cause unanticipated interactions between the
	components, especially between the digital and physical parts, leading to
	global emergent behaviour that could threaten safe and reliable system
	operation.

	The uncertainty associated with a CPS, which is reinforced by the
	insufficient knowledge and understanding about the system and its dynamics, can
	be attributed to three primary complexity drivers~\cite{complexity}.
	\textit{Structural complexity} characterises the underlying CPS infrastructure
	and architecture, as these systems have numerous components with convoluted
	interactions.
	\textit{Dynamic complexity} arises due to the difficulty to understand the
	system behaviour and system dynamics entirely.
	\textit{Organisational complexity} arises because of the heterogeneity of CPSs
	and the adopted engineering processes.
	These complexity drivers are certainly interdependent and are manifested
	both at design time and while the CPS is in operation.

	Safety assurance methods are typically employed as a means of carrying out a
	planned and systematic assessment aiming to tackle some uncertainty facets and
	provide sufficient confidence that the CPS achieves tolerable safety
	levels~\cite{safety}.
	These methods can be classified into three main categories: hazard
	identification and analysis (e.g., preliminary hazard analysis, failure mode
	and effects analysis); risk analysis and evaluation (e.g., reliability block
	diagrams, Markov models); and verification and validation (e.g., model
	checking, testing, runtime verification). Depending on the application domain,
	the set of acceptable assurance methods is provided in relevant safety
	standards (e.g., ISO26262 in automotive, ARP4761 in aviation).
	The outcome from these assurance methods drives the derivation of safety
	requirements and their subsequent refinement and allocation to architectural
	and software artifacts the span the CPS engineering stages. The recursive
	refinement and allocation process enables to formulate an
	\textit{incomplete} assurance case and reason about the system safety
	through using evidence produced from using these assurance methods across the
	CPS lifecycle~\cite{safetycases}.

	Other uncertainty facets, however, including autonomous decision-making enabled
	by artificial intelligence algorithms, cannot be easily anticipated at design
	time as they are primarily manifested while the CPS is in operation.
	This unique characteristic of CPSs increases the unpredictability of their
	behaviour and can lead to a global emergent CPS behaviour putting safe and
	reliable system operation in jeopardy.
	Since CPSs are typically deployed in safety-critical application domains, this
	emergent behaviour can cause accidents with catastrophic outcomes.
	Existing safety assurance methods (including those recommended by
	safety standards like ISO26262 and ARP4761) do not have the ability to capture
	and establish these uncertainty types.
	Thus, there are limited means for deriving quantifiable requirements that will
	drive the design and implementation of software underpinning the autonomous
	decision-making processes of CPSs and the extraction of traces that enable to
	validate the compliance of the implemented software with its requirements.

	We argue that there is a need for developing new or adapting existing safety
	assurance methods, and devising suitable abstractions that enable
	software-intensive CPSs to establish a rationale for the decisions being made
	and to link these decisions with safety requirements.
	This might take the form of advanced hazard identification techniques
	or new testing and verification techniques.
	Achieving this objective will provide the technical foundations for
	deriving explainable and assured software, thus increasing our confidence for
	trustworthy CPS behaviour and providing the means to produce fully-fledged CPS
	assurance cases~\cite{safetycases}.

\end{conf-abstract}
\newpage

\begin{conf-abstract}[]
	{Explainability of Cyber-Physical Security Threats}
	{Christopher Gerking}
	{Software Engineering Research Group\\Paderborn University, Paderborn, Germany\\ {christopher.gerking@upb.de}}
	\indexauthors{Gerking!Christopher}
\noindent
	\subsection*{Explaining Secure Design Decisions}

	Developing software requires engineers to ensure consistency between the
	\emph{design} of a system and the quality \emph{goals} of the system's
	stakeholders. Such goals ultimately determine the design of a system, and the
	design decisions made by engineers affect the degree to which goals are
	achieved. In this regard, \emph{security} is a quality property that is special
	as it involves \emph{threats} as an additional dimension beside design and
	goals~\cite{gerking:Tur17}. A threat represents the source of a security incident, and thereby
	relates to both goals and design: security goals must cover a
	reasonable amount of threats in order to be consistent, whereas threats may be
	mitigated or even facilitated by particular design decisions.

	However, design decisions that are made to mitigate threats are often hard to
	explain. The reason is that such mitigations are no longer clearly traceable to
	security goals. Instead, they are motivated in terms of unimaginable and
	seemingly unreal threats. Due to the vague and fast-moving nature of threats,
	explaining the harm they cause is often only possible in a post-incident way,
	i.e., if a threat has already materialized in form of a security incident. The
	intention to prevent such incidents beforehand is known as \emph{security by
		design}, but the successful application of this principle suffers highly from
	the aforementioned problems in terms of explainability. Systems with
	unexplainable mitigations may either be used or operated in an insecure way, or
	fully rejected by users. In the worst case, unexplainable mitigations may even
	be neglected by engineers, thereby introducing serious security vulnerabilities.

	Engineers may encounter this problem not only when explaining mitigations to
	stakeholders, but also when it comes to the explainability of results produced
	by automated security analyses. In particular, analyzable security properties
	like \emph{noninterference}~\cite{gerking:GM82} are known to be \emph{hyperproperties}
	which are properties of sets of execution traces~\cite{gerking:CS08}. Therefore, a
	counterexample that indicates a security vulnerability refers to more than one
	trace. Thus, it is hard to map such a counterexample to a real-life security
	incident with a clear indication of harm, thereby affecting the explainability
	of the analysis results.

	\subsection*{Challenging Characteristics of Cyber-Physical Systems}

	The nature of cyber-physical systems implies several characteristics that
	further intensify the need to explain security threats. First of all,
	cyber-physical systems integrate artifacts from multiple engineering
	disciplines. Accordingly, security threats such as \emph{side channels} exploit
	physical effects. Therefore, detecting and mitigating such threats involves
	disciplines other than software engineering and requires a discipline-spanning
	explanation at the level of \emph{systems engineering}.

	Second, the physical environment of systems imposes hard real-time constraints
	on their behavior. However, this real-time behavior must not enable attackers to
	infer sensitive information indirectly from the response times of a system.
	Such leaks, called \emph{timing channels}, must be detected by corresponding
	analyses. Therefore, the time aspect needs to be taken into account during the
	explanation of threats as well.

	Third, cyber-physical systems are increasingly self-adaptive by reconfiguring
	their structure or behavior according to a certain situational context. Such
	adaptations must not compromise the security of a system, which is why security
	considerations need to be taken into account during the analysis and planning of
	adaptations. Thereby, a system might even adapt its level of protection to a
	specific security situation. However, when it comes to the explainability of
	such adaptations, the corresponding security threats need to be integrated into
	the explanations as well.

	\subsection*{Research Agenda}

	Model-driven engineering is frequently applied to the development of cyber-physical systems that
	need to be \emph{secure by design}. In previous work, we proposed a formal, model-driven security analysis that takes the real-time behavior of systems into account~\cite{gerking:GSB18}. In this respect, the counterexamples obtained from the analysis suffer from a missing link to concrete, explainable security threats. To improve the explainability of analysis results, additional work
	is needed in order to link counterexamples to explainable threat \emph{scenarios}. Furthermore, an additional effort is required to address the challenge of executing and explaining self-adaptations in a way that takes security threats into account.



\end{conf-abstract}
\newpage
\begin{conf-abstract}[]
	{Automated Experimentation for Online Learning and Adaptation}
	{Ilias Gerostathopoulos}
	{Software and Systems Engineering\\Technical University Munich, Germany\\ {gerostat@in.tum.de}}
	\indexauthors{Gerostathopoulos!Ilias}
\noindent
	\subsection*{ES4CPS problem}
	
	Cyber-physical systems (CPS) are typically large systems composed of a number of independently developed and tested components that may opportunistically collaborate with each other. 
	Consider, for example, case of several robots collaborating in pushing a door open or passing through a narrow passage.
	As another example, consider the complex interactions of several vehicles in a traffic network (e.g. a city or a highway) that need to collaborate in order to reduce trip times, fuel consumption and CO2 emissions. 
	One problem in large CPS is that classical engineering approaches for developing complex systems that follow the specify-develop-test-analyze-deploy loop may not apply.  
	A particular problem exists in the analyze phase that, traditionally, precedes the deployment phase: complex large CPS are difficult to model and analyze before deploying them in production environments, where the whole range of interactions (including emergent ones) is manifested. 
	At the same time, complex CPS need to be explainable and predictable in their behavior. 
	To make things worse, they may need to adapt at runtime to deal with changes in their runtime context (e.g. in the robotic case, consider an extra obstacle that makes the door harder to open).
	
	A particular problem I am interested in is that of optimizing a traffic router used by a number of cars in a city (published as ``CrowdNav'' exemplar at SEAMS 2017\footnote{S. Schmid, I. Gerostathopoulos, C. Prehofer, and T. Bures, “Self-Adaptation Based on Big Data Analytics: A Model Problem and Tool,” in Proc. of SEAMS 2017, \url{https://github.com/iliasger/CrowdNav}}). 
	The problem consists of selecting the values for a number of (numeric) parameters or a router used by all cars in a city in order to minimize the trip time of the cars. 
	The complexity lies in that the cars have random destinations and have complex interactions in the physical world resulting in traffic slowdowns, jams and accidents. 
	Moreover, the number of cars in the city may increase or decrease following reality in which traffic the varies within a single day. 
	One question is how to find a solution to the above optimization problem in the fastest way, since the environment may change before the optimization finishes. 
	Another challenge is how to ensure the stability of the solution.
	A third challenge lies in handling complaints of drivers when the optimization algorithm tries risky configurations. 
	
	\subsection*{ES4CPS solution}
	
	With respect to CrowdNav, we have so far applied a combination of factorial analysis and Bayesian optimization in order to optimize trip durations with a very small number of tries\footnote{I. Gerostathopoulos, C. Prehofer, and T. Bures, “Adapting a System with Noisy Outputs with Statistical Guarantees,” in Proc. of SEAMS 2018, 2018, pp. 58–68}.  
	In this solution, the traffic system together with the router is viewed as a black box with the seven parameters of the router as input and the trip duration of cars as a single output. 
	First, a factorial design is applied in order to examine the effect of different, pre-defined, combinations of parameter values to the output. 
	Then, only a subset of these parameters, the ones with the stronger, statistically significant, effect on the output, are selected to be optimized via Bayesian optimization. 
	Bayesian optimization essentially performs a search in the space of input parameters in order to learn which combination minimizes the value of the output. 
	The current solution does not take into account changes in the environment that may lead to new optimization rounds. 
	Moreover, it does not cater for reusing knowledge about the effect of inputs to outputs across different system situations (e.g. low traffic, medium traffic, high traffic).
	These are two highly interesting topics that can be investigated further. 
	
	Broadening the scope, I envision a method that automates the experimentation with a system in production by learning system models from data (in the case of CrowdNav, important data are system outputs, i.e. trip durations) and using them for adapting to runtime changes.
	In order to be used in real-life settings, the method needs to provide guarantees w.r.t. the use of the learned models and support rump-up and abortion of experiments, similar to A/B testing, to deal with experiments that have high cost. So far we have investigated the use of statistical methods such as t-test and factorial ANOVA, to provide statistical guarantees\footnote{As above}. 
	
	\subsection*{Contributing Expertise}
	
	I have working knowledge in statistical testing and Bayesian optimization. 
	I have expert knowledge of the CrowdNav testbed which I have co-developed, maintained, and extended. 
	Finally, I have co-developed a tool that allows one to specify the inputs and the outputs of a system-to-be-optimized and run different experimentation pipelines on it\footnote{\url{https://github.com/iliasger/OEDA}}.
	
	\subsection*{External Expertise needed}
	
	Expertise in machine learning (and in particular transfer learning), data science, statistical methods, optimization, and search methods would be very useful in tackling some of the hard challenges of automated experimentation. 
	As another interesting topic, expertise in requirements engineering would be useful in determining the optimization goals and criteria and the configuration points in a system-to-be-optimized.   
	
	
\end{conf-abstract}
\newpage
\begin{conf-abstract}[]
	{Towards Explainable Controller Synthesis and\\Reinforcement Learning for Cyber-Physical Systems}
	{Joel Greenyer}
	{Software Engineering Group,\\Leibniz Universität Hannover, Germany\\{greenyer@inf.uni-hannover.de}}
	\indexauthors{Greenyer!Joel}
	\noindent

\subsection*{Future CPS and Understanding their Behavior}

In domains like transportation and automation, Cyber-Physical Systems (CPS) control more and more complex and critical tasks. The complexity of the controlled processes, user interactions, the distributed nature of the systems, and the uncertainty in the environments are just some of the factors that make CPS development increasingly challenging. This complexity must be met with more high-level development techniques that raise the abstraction level from imperative programming or state-based modeling towards modeling specifications or goals on a level that corresponds more closely to how humans conceive requirements; this could be scenarios~\cite{Greenyer2017a}, temporal logic specifications, or controlled natural languages; controller synthesis and reinforcement learning techniques will then construct or learn the executable software that exhibits the desired behavior. However, when the system behavior is constructed algorithmically, and even changes at run-time, this creates new challenges for comprehending the system behavior---at design-time and at run-time.

From our experience with game-based controller synthesis (CS) from scenario-based specifications \cite{Cordy2015,Brenner2015,Greenyer2017a}, we found that it can be very difficult to understand specification inconsistencies when CS is unable to construct an implementation (valid control strategy). It can also be that issues remain in the specification, but CS is still able to construct an implementation (typically due to over-optimistic assumptions).

In reinforcement learning (RL), there is a similar challenge: Very often one does not know whether the system can enter states where it will struggle or fail to reach its goals, or whether it may choose a strategy (called \emph{policy} in RL) that is only locally optimal (for example a robot that always cleans some dirt in the kitchen instead of cleaning the whole apartment).

\subsection*{Explainability at Design-Time and Run-Time}

We ultimately desire systems that are able to explain their behavior, misbehavior, struggles, or other aspects of interest---at design-time and run-time. But how could such (self-)explanation capabilities look like? First of all, the question is what kinds of questions human stakeholders need answers to? From our experience, some examples are \enquote{Can you (the system) reach a certain state from another / the current state?} (Reachability), \enquote{What action(s) will you perform under certain conditions?}, \enquote{Which goals/requirements cause you to choose these actions in a certain state?}, \enquote{What is your goal in a certain state?}, \enquote{Under what conditions do you perform certain actions?}, \enquote{When are you struggling to achieve a goal?}, or \enquote{How can the environment/user help the system to achieve a certain goal?}

Such questions must be expressible in a structured query language, and we require an engine that computes answers to these queries. On what basis should the responses be computed? They should be computed, first of all, based on a synthesized or learned strategy, but also based on the specification or goal models, or other associated design artifacts. At run-time, also a log of past behavior can be considered. The answers must be as concise as possible. For example, queries that result in sets of states should summarize the defining features of the states in the set. Questions that result in (sets of) traces, cycles, or (sub-)strategies should also summarize their defining features, highlight relevant events, commonalities, or key decision points. Eventually, human-friendly front-ends are required that allow users to refine queries, or interactively simulate certain behaviors of the system.

\subsection*{Research Directions}

Achieving the above requires research in multiple directions: (1) on algorithms for finding and filtering certain aspects from traces or (counter-)strategies. Some work already exist in this direction, e.g., \cite{Kuvent:2017:SJV:3106237.3106240}; (2) on techniques for rendering such behavioral aspects as higher-level models, possibly scenarios; (3) on  program/specification repair \cite{Greenyer2017e}.

Extending these ideas in the direction of RL is interesting. Usually learned policies in RL are represented as state-action matrices, or various abstraction techniques are employed, including neural-networks that classify states. One question is whether higher-level, explanation models can be learned, or initial design models be refined, in the learning process, so that answers to queries as described above can be computed. Where states are classified using neural networks, techniques would be necessary to extract their defining features in a comprehensible form. Reinforcement learning is successfully applied where no comprehensive model of the system's behavior or environment exist (model-free learning), but there exist approaches that combine model-free learning and model-based learning or planning (e.g. the dyna principle,~\cite{Sutton1991}) for increased efficiency---we could similarly combine model-free learning with the learning of explanation models, which could then serve to answer queries as above at run-time.

\end{conf-abstract}\newpage
\begin{conf-abstract}[]
	{Improving the Explainability of Autonomous and \\Intelligent Systems with Conversational Interfaces}
	{Nico Hochgeschwender}
	{German Aerospace Center (DLR), Germany\\{Nico.Hochgeschwender@dlr.de}}
	\indexauthors{Hochgeschwender!Nico}
	\noindent
	
\subsection*{Motivation}
Autonomous and intelligent systems such as self-driving cars, smart grids, advanced satellites or service robots are to a large extent \emph{software-intensive systems} where increasingly advanced functionality is provided by novel learning, reasoning, planning and decision-making abilities. Those abilities are not only exploited by autonomous systems to achieve non-trivial goals, but also to autonomously adapt their course of action to the various and changing requirements induced, for example, by the systems task, its complex, cyber-physical environment or by varying user requests.   
Even though autonomous and intelligent systems are successfully deployed in real-world environments they lack the ability to communicate and explain \emph{why} and \emph{how} they have made particular decisions. Therefore, we argue that every autonomous system needs to be \emph{explainable}. To this end, an explainable system should be -- at any point in time -- capable to express and explain its internal decisions, functional abilities and limitations, and its past, current and potential future behavior to different stakeholders (e.g. developers, users, lawyers or certification agencies). It is important to emphasize that those explanations need to take into account the full software stack of autonomous systems which poses, among others, several representational challenges for platforms, frameworks and infrastructures as one needs to investigate which knowledge is required for the explanations. 

\subsection*{User-centered Approach}
Implementing explainable systems is not only crucial for users to develop trust, but also of paramount importance for developers, insures and lawyers to find out what and why something went wrong in the case of, for example, an accident or failure. In order to deal with the varying explainability requirements imposed by different stakeholders we propose a user-centered approach to design and develop advanced \emph{user interfaces}. Thereby, our goal is to develop interfaces which are capable to satisfy stakeholder needs by providing configurable means for communication and explanation which make use of concepts and terms from the stakeholders domain. 
\subsection*{Conversational Interfaces}
To this end, we propose to exploit conversational interfaces (also known as chatbots) as a technology to improve the explainability of autonomous systems. Conversational interfaces offer a great opportunity to communicate with autonomous and intelligent systems in a convenient way using human languages. The idea that chatbots can contribute to improving the transparency of autonomous systems is supported by our recent work described in~\cite{Jentzsch2019} where we demonstrated how a conversational 
interface can enhance the task of exploring and understanding the abilities of an autonomous robotic agent. However, to fully exploit the advantages of conversational interfaces (e.g. accessibility) for improving the explainability of autonomous systems we need to better understand when, how and why users interact with chatbots for the sake of seeking explanations. We argue that conversational analysis techniques could serve as a tool 
to investigate and assess those requirements. For example, conversation analysis describes the interaction with a chatbot as sequentially organized actions. The different actions and interactional practices such as questioning, devices (e.g. upper vs. lower case writing) and turn formats (combination of practices and devices) need to be properly investigated from an explanation point of view. Persons seeking explanations usually show their lack of information by asking questions. Therefore, a conversational interface should be capable not only to deal with different question types (e.g., known-answer and unknown-answer questions), but also to deal with varying epistemic stances (expectations of the knowledge of the other speaker). This is in particular important if a chatbot should be used by different stakeholders with different background knowledge and possibly varying concepts and terms. We argue that the above mentioned considerations are important for developers to design useful conversational interfaces which go beyond the intent and task-based conversational interfaces currently available.

\end{conf-abstract}
\newpage
\begin{conf-abstract}[label]
	{Formal-methods based motion-planning for mobility-on-demand and its application on fleets of autonomous vehicles -- an ES4CPS problem}
	{Jesper	Karlsson}
	{Robotics Perception and Learning \\KTH Royal Institute of Technology, Sweden\\jeskarl@kth.se}
	\indexauthors{Karlsson!Jesper}
	\noindent

\subsection*{The ES4CPS problem}

My research is focused mainly on the development of efficient (i.e., compositional and/or decentralized) planning, scheduling and control techniques for fleets of autonomous vehicles with the use of formal methods. Recent developments in the area involve bottom-up decentralized control for robotic teams from possibly dependent local linear temporal logic formulas as well as decomposition of global specifications into individual, independent formulas for individual agents. My work puts emphasis on state-space and knowledge representation, reasoning, and high-level decision making aspects. Specifically, logical reasoning about the other agents’ states, knowledge and plans. The general goal is to develop a framework enabling efficient use of rich, yet user friendly, temporal logic as a specification means for complex goals in collaborative or competitive multi-vehicle environments. Some of my recent work includes extensions of temporal logic towards handling explicit time and probabilistic bounds via investigating timed and probabilistic versions of temporal logics and hybrid and Markovian models of the system components. Targeted application areas involve routing, scheduling and control of autonomous vehicles for mobility on demand. We focus on sampling-based approaches of motion planning for these applications. The ES4CPS problem that has been my main interest as of late is that of human-robot interaction for use in autonomous vehicle scenarios and how to model human behaviour so as to provide provable guarantees on safety and performance of the underlying CPS.

\subsection*{My Expertise}

My main expertise lies in motion planning for autonomous robots (specifically autonomous
vehicles) with temporal logic specifications and quantitative guarantees. As stated in previous section, the focus of my work has been to develop a framework for specifying
complex spatial-temporal goals using temporal logic, with the main application being
mobility-on-demand for autonomous vehicles.
I have studied both the high-level routing problem as well as the motion-planning problem
for autonomous vehicles. Recent publications as a result of my work is to ECC~\cite{karlsson:KT18} as
well as ICRA~\cite{karlsson:Ka18}. Recent efforts have been focused mainly on extending our work into
handling probabilistic and stochastic behaviour in complex (dynamic) environments.The
first step of doing this is formalizing an intention-aware motion planner. I believe that
Intention prediction is an extremely important piece in motion planning for fleets of
autonomous vehicles acting in complex and dynamic environments.

\subsection*{External Expertise}

Given that my recent focus has shifted into probabilistic and stochastic knowledge repre-
sentation as well as intention-aware motion-planning in environments with autonomous
vehicles and pedestrians, an important piece that is missing is the human-robot interaction.
For instance, modelling humans, in order to predict behaviour, is a very complex problem
that has little concrete published work in the literature. External expertise that I would
benefit greatly from is therefore anyone who is currently, or has in the past, worked with
modelling human behaviour or any form of human-robot interaction. I am also interested
in the combination of formal methods and reinforcement learning in order to provide
quantitative guarantees on the resulting policy.

\subsection*{Summary}

Since my main research interest lies in motion planning for autonomous robots with
temporal logic specifications, using Formal-methods based approaches. This is the expertise
that I can contribute with during the course of the seminar. As current research into the
probabilistic versions of the same problem, as well as the human-robot interaction that is
needed for these approaches, I am very interested in exchanging ideas regarding human-
robot interaction and reinforcement learning combined with formal methods.

\end{conf-abstract}\newpage
\begin{conf-abstract}[]
	{Security Explainability Challenges in Cyber-Physical Systems}
	{Narges Khakpour}
	{Linnaeus University, Sweden\\ {narges.khakpour@lnu.se}}
	\indexauthors{Khakpour!Narges}
	\noindent
	The growth in development and deployment of Cyber-Physical Systems (CPS) has lead to emerging new security problems and challenges.
	A CPS usually consists of various types of software and hardware components, possibly provided by different vendors, that interact using different communication technologies.
	A CPS can be vulnerable to attacks that target its cyber elements, its physical parts, its communication link or a combination of them.
	Attacks to single components or single layers are usually easier to detect and mitigate, while attacks raised and amplified by the inter-dependence and interactions between cyber and physical components, that are usually provided by different vendors, are harder to detect, explain and protect against.
	In particular, coordinated attacks that target multiple components in a CPS and exploit different vulnerabilities in cyber and physical layers are very challenging to identify, explain, asses and mitigate.
	For instance, in an attack against Ukrainian electricity grid in 2015  that caused a power outage which hit approximately 225, 000 people~\footnote{\url{ https://ics-cert.us-cert.gov/alerts/IR-ALERT-H-16-056-01}}, the phone network was blocked in parallel which made it very hard for the technicians to communicate with each other to fix the issue.

	We believe that \emph{explainability} can provide a suitable means to design and develop more trusted and secure systems. \emph{Security explainability} refers to the ability to understand, trust and manage a system and is defined as a new paradigm to manage security in~\cite{khakpour:ViganMagazzeni2018}.
	The characteristics of CPS, specially their complexity and components heterogeneity,
	demand new techniques to design,  explain, analyze and enforce security that consider such specific features.
	We point to a few challenges that should be addressed to realize explainable security in CPS.

	\textbf{Security Explainability by Design} Many CPSs are not usually designed with security in mind~\cite{khakpour:CPSSecuritySurvey2017}, or the focus is usually on security of individual components.
	We need new design methods for CPS in which security is considered from the beginning of the system life-cycle, consider the specific security requirements of CPS and develope the software in a way that one can detect and troublshoot malicious behaviors or security weaknesses in future. One approach to tackle this challenge is to use domain-driven security~\cite{khakpour:Basin2011ADO} that aims to facilitate secure software development by integrating
	security requirements with software development processes.

	The model-driven security methods express security requirements in addition to the system design, propose analysis techniques
	to reason about the system security, and generate the code automatically from the models to enforce  security
	policies.
	Such design techniques, tailored to CPS, should (i) take into account interactions of the physical and cyber layers and their potential security implications, (ii) provide proper technical documentations to discover, explain and fix security violations in future, e.g. the (verified) security implications of a change in a component in interaction with other components of the eco-system, or components' compatibility information, (iii) use advanced logging mechanisms and formats that enable us extract helpful information about the system behavior in case of a suspicious behavior to explain and root-cause analyze the undesired behaviors and (iv) offer secure logging mechanisms that don't lead to confidential information leakage.

	\textbf{Evolving Security Mechanisms}
	Today's CPS are designed in a way to get evolved at runtime and adapt themselves to the changing requirements and the operational environment, e.g. by applying a security patch, and as such, any change in the system can introduce new vulnerabilities and make its protection mechanisms ineffective.
	Hence, with an increasing number of attacks and the evolution and adaptation of systems, the protection mechanisms must subsequently evolve and be improved over time to face future attacks. Reactive security techniques (like encryption, Intrusion Detection Systems (IDSs) etc), although very useful, can no longer be solely effective in such dynamic environments, and developing proactive and adaptive mechanisms is becoming an indispensable need.

	To realize effective self-protecting systems, the security implications of any change should be checked before application, and each system component must be provided with (verified) security signatures that explain the potential vulnerabilities, security threats and their implications.
	This will allow us to explain any undesired behavior in the eco-system by analyzing components' security signatures and their interactions. This will let us evaluate and explain the security impacts of an adaptation too.


\end{conf-abstract}
\newpage
\begin{conf-abstract}[]
	{Explainable Self-Learning Self-Adaptive Systems}
	{Verena Klös}
	{Softw. and Embedded Syst. Eng., TU Berlin, Berlin, Germany\\ {verena.kloes@tu-berlin.de}}
	\indexauthors{Klös!Verena}
	\noindent
	\subsection*{Self-Learning Self-Adaptive Systems}
	%
	Self-adaptivity enables flexible solutions in dynamically changing environments. However, due to the increasing complexity, uncertainty, and topology changes in cyber-physical systems (CPS), static adaptation mechanisms are insufficient as they do not always achieve appropriate effects. Furthermore, CPS are used in safety-critical domains, which requires them and their autonomous adaptations to be safe and, especially in case of self-learning systems, to be explainable.
	To achieve this, we have developed an engineering and analysis approach for self-learning self-adaptive
	systems~\cite{kloes:Kloes2016, kloes:Kloes2018a, kloes:Kloes2018b} based on our notion of timed adaptation rules. These rules are extended condition-action rules, that additionally include a timed expected effect of the adaptation actions.
	In contrast to techniques like online planning, timed adaptation rules make adaptation decisions explicit and comprehensible due to their explicit application condition and timed expected effect.

	At run-time, we employ history information of system and environment parameters together with history information of applied adaptation rules to retrace past adaptation decisions and record whether their expected effect was achieved and in which system contexts. We use this information to evaluate the accuracy of adaptation rules and to dynamically improve the adaptation logic through learning. Our rule accuracy evaluation detects inaccurate adaptation rules and classifies the observed effects of previous rule executions w.r.t. the observed deviation from the expected effect. The evaluation results are used to refine deviating adaptation rules into more specific context-dependent rules that capture the observed effects. This enables the accurate co-evolution of the adaptation rules.
	To learn adaptation rules for new situations that are not covered by the existing rules, we employ executable run-time models of the system and its environment. Those models provide an interface to update them with the dynamically gathered knowledge of the system and environment to realise co-evolution of models and actual environment.
	As a result, we achieve an updated adaptation logic that consists of more accurate, yet comprehensible, timed adaptation rules.
	A separate verification phase enables us to provide offline and
	online guarantees of evolving adaptation logics based on human-comprehensible formal models.

	\subsection*{Traceability of Decisions for Comprehensibility}
	\label{sec:traceability}

	To achieve explainability of autonomous adaptation decisions, we propose precise retracing of adaptation decisions, containing their trigger (i.e., violated adaptation goals),  the context (i.e., the current values of system and environment parameters), the chosen rule, the expected, and the actual effect (in terms of observable changes
	and goal satisfaction). Based on that, we can generate answers to the following questions: ``Why was adaptation necessary?'', ``Why did you choose this option?'', ``Which assumptions were made?'' and ``Which results were achieved?''. With further processing of different traces that lead to the same adaptations, it might  also be possible to detect the cause of an adaptation which might be more interesting than the trigger.
	To explain our adaptation rule learning, we provide  an explanation basis for the following questions: ``Why was learning necessary?'', ``How were the results achieved?'', ``Which assumptions were made?'' and ``Which results were achieved?''.
	To this end, we build explanation objects that contain the original adaptation rule (if existent), the kind of learning (observation- or simulation-based learning), the learning result (refined context-specific rules, or newly learned rule), and, for observation-based learning, the underlying evaluation results.
	To explain the results of simulation-based learning, we also log the updated run-time models as these will further evolve over time. Furthermore, we add the fitness function of the learning algorithm to explain the evaluation of found solutions.

	\subsection*{From Comprehensibility/Traceability to Explainability}
	Each object in our explanation basis can already be seen as explanation for a single adaptation/ learning decision. While this explanation format is sufficiently comprehensible for experts, i.e. self-adaptive software engineers, further processing to generate textual explanations may be beneficial for non-experts and is part of future work. Furthermore, we plan to consider customized explanations for different target groups. This can be achieved by generating customized textual explanations based on our explanation base.  Currently, our explanation base contains all possibly relevant information. To avoid information overload, but still provide sufficient information, we aim at providing filtered information to answer questions, and to cooperate with cognitive science to investigate human needs and expectations on explanations. Additionally, we plan to investigate the use of machine learning on user feedback to learn the characteristics of helpful explanations.


\end{conf-abstract}
\newpage
\begin{conf-abstract}[label]
	{Explainable Automotive Emissions}
	{Maximilian Köhl}
	{Saarland University, Saarbrücken, Germany \\mkoehl@cs.uni-saarland.de}
	\indexauthors{Köhl!Maximilian}
	\noindent
    In modern cars, embedded software controls the combustion and exhaust purification process. A multitude of parameters is being influenced so as to lower the emissions of toxic exhaust gases. In order to ensure that these emissions do indeed stay within reasonable limits, cars need to pass certain test procedures on chassis dynamometers for type approval. Until recently, the test setup and conditions were specified precisely, for reproducibility and comparability reasons.  In what has been known as the diesel emissions scandal, \emph{doped} \cite{koehl:PLS:2017:SDOPE} control software has been put in place to detect running emission tests and alter the combustion and purification process to manipulate the test results effectively rendering the results meaningless.
     
    In response, the European Union has defined a new test procedure to test for \emph{Real Driving Emissions} (RDE) \cite{koehl:LEX:32016R0427}. The RDE is designed to provide insights into the actual in-the-wild emissions of vehicles.
    With the emerge of \emph{Portable Emissions Measurement Systems} (PEMS) which can be carried with a vehicle under scrutiny, it is now possible to gather these insights, also in combination with other relevant parameters influencing emissions.
    
    This opens a research agenda from emission monitoring to emission analysis, and eventually emission explanation. 
    Making exhaust emissions explainable will enable answers to questions like: Does a vehicle emit increased levels of certain gases because of the driving behavior, a manipulated purification or combustion process, or a fault in the system? 
    
    In recent work, we formalized the RDE test procedure using the stream-based specification language Lola \cite{koehl:RV:RDE}. Based on this formalization a runtime monitor was synthesized which we used to conduct RDE-like tests based on low-cost equipment we developed. The results are promising. We were able to gather valuable insights into the exhaust emissions and purification control software of an Audi A7. Our results suggest that our particular car contains an illegal defeat device manipulating the exhaust purification process under certain circumstances \cite{koehl:LPAR-22:Verification_Testing_and_Runtime}, a suspicion which recently has been confirmed by Audi.
    
    \paragraph{What is an ES4CPS problem that I am interested in?} 
    The diesel emissions scandal has demonstrated that opaque control software can easily be used against the interest of society.
    Based on our earlier success with monitoring the in-the-wild behavior of an Audi A7, I would like to develop means to understand the emission behavior of vehicles in-the-wild. From a broader perspective, I would like to enable laypersons with low-cost equipment to get valuable insights into complex CPS. To this end, I am interested in explanatory models for systems controlling chemical processes eventually allowing us to explain and predict automotive emissions based on driving behavior and external circumstances.
    
    \paragraph{What is the ES4CPS-related expertise that I can contribute?} Related to the aforementioned problem, I have expertise based on my previous work in how to gather insights in automotive emissions, both, from a theoretical and practical perspective.
    
    \paragraph{What external expertise do I need?} Considering the theoretical part, I could use external expertise on how to faithfully model the software controlled chemical processes found in modern cars. Considering the practical part, I could use expertise on how extract information from embedded car controllers using non-standardized interfaces.

\end{conf-abstract}

\newpage
\begin{conf-abstract}[]
	{Model-based Testing Strategies for Configurable Software with Unbounded Parametric Real-Time Constraints}
	{Lars Luthmann and Malte Lochau}
	{Real-Time Systems Lab, TU Darmstadt, Darmstadt, Germany\\ {lars.luthmann@es.tu-darmstadt.de}, {malte.lochau@es.tu-darmstadt.de}}
	\indexauthors{Luthmann!Lars, Lochau!Malte}
	\noindent
		Non-functional requirements like real-time behaviors are of ever-growing interest in many application domains of \emph{cyber-physical systems (CPS)}.
		Additionally, we require CPS to be \emph{configurable} such that variability of systems is taken into account.
		Consequently, existing modeling formalisms and analysis techniques for reasoning about time-critical behaviors have to be adapted to engineering of configurable CPS, too.
		Featured timed automata (FTA)~\cite{luthmann:Cordy2012} extend \emph{timed automata (TA)}~\cite{luthmann:Alur1990}, a modeling formalism for discrete event/continuous time software systems, with annotation-based variability modeling~\cite{luthmann:Czarnecki2005} and (symbolic) family-based model-checking~\cite{luthmann:Classen2011} of configurable CPS.
		Here, we present configurable parametric timed automata (CoPTA)~\cite{luthmann:Luthmann2017} to further extend expressiveness of FTA by freely configurable and a-priori unbounded timing intervals of real-time constraints known from parametric timed automata (PTA)~\cite{luthmann:Henzinger1994}.
		Hence, CoPTA models impose infinite configuration spaces, making variant-by-variant analysis practically infeasible.

		Instead, we present a family-based test-suite generation methodology for CoPTA models ensuring symbolic location coverage for every model configuration.
		Furthermore, we define a novel coverage criterion, called Minimum/Maximum Delay (M/MD) coverage~\cite{luthmann:Luthmann2019}, requiring every location in a CoPTA model to be reached by test cases with minimum/maximum possible durations, for systematically investigating best-case/worst-case execution times.
		Consequently, we extend our family-based test-suite generation methodology to also achieve M/MD coverage on CoPTA models.
		Our evaluation results, obtained from applying our CoPTA tool to a collection of systems, reveal efficiency improvements of test-suite generation for configurable CPS, as compared to a variant-by-variant strategy in case of finite configuration spaces.
		Concerning the goals of the Dagstuhl Seminar on ES4CPS, we consider the following.

		\paragraph{What is an ES4CPS problem, and/or what is an ES4CPS solution, that we are interested in?}
		Based on our testing methodology for configurable real-time CPS, our techniques may be used to systematically inspect configurable CPS to get a better understanding of the impact of parameters on execution times.
		For instance, we may explain best-case/worst-case execution times of CPS.
		Furthermore, one needs to first find a suitable model to express interesting aspects of explainable software to tackle these problems, whereupon a notion of testing may be defined.

		\paragraph{Insights from the Workshop on ES4CPS.}
		Concerning testing, explainability is especially interesting in the context of faults (i.e., failed test-case executions)~\cite{luthmann:Halpern2005causes,luthmann:Halpern2005explanations}.
		Here, one goal may be to find the \emph{cause} (or all possible causes) of a fault to explain erroneous behaviors.
		In order to obtain a cause, we need to isolate a minimal trace leading to the fault~\cite{luthmann:Groce2003}.
		Further challenges arise when considering the oracle problem, i.e., evaluating and explaining (un-)expected behaviors of test executions.
		Moreover, proper traceability information between artifacts and tests is pivotal for testing and the explainability of (un-)expected test results.

\end{conf-abstract}
\newpage

\begin{conf-abstract}[]
	{Strategy Patterns for Multi-Agent CPS}
	{Claudio Menghi}
	{University of Luxembourg, Luxembourg\\claudio.menghi@uni.lu}
	\indexauthors{Menghi!Claudio}
	\noindent
	\subsection*{Introduction}
	CPS are software systems that are tightly integrated with their environments.
	These systems usually rely on a set of  agents that collaborate or compete for achieving certain goals.
	Defining how these agents should behave, i.e., their missions,  is complex and error-prone~\cite{menghi:Menghi:2018:PSP:3183440.3195044} and  plays a critical role in the definition of  explainable CPS.
	Explainability refers to the capability of  documenting and defining the mission the agents must achieve  in a human-comprehensible and machine-processable format;
	two requirements that are apparently in conflict.
	On the one hand, the mission specification should be \emph{machine-processable}, i.e., it must be precise and rigorous.
	This requirement intrinsically calls for the use of mathematical instruments, such as logical languages, that precisely define the semantics of missions.
	However, practitioners are often unfamiliar with the intricate syntax and semantics of logical languages which make  these languages inappropriate for practical adoption~\cite{menghi:dwyer1999patterns}.
	On the other hand, the mission specification  should be \emph{human-comprehensible}, i.e.,
	it should be easily managed and defined by people with low technical expertise.
	However, comprehensible mission specifications, such as the one defined by using natural language, are often not machine-processable.
	Even if specification patterns tailored for the CSP domain have been recently  proposed as a solution for this problem~\cite{menghi:Menghi:2018:PSP:3183440.3195044,menghi:PSALM}\footnote{www.roboticpatterns.com}, these patterns are defined for single-agent CPS applications, and thus they do not explicitly address the multi-agent case.

	In multi-agents CPS applications, \emph{strategies} play a pivotal role.
	Strategies are  plans of actions designed to achieve the desired goal and specify when and how the different agents should perform their actions.
	While 
	machine-processable strategy aware logics   have been proposed in the literature, such as Strategy Logic~\cite{menghi:chatterjee2010strategy},
	they are not used in practice since they are not easily comprehensible by non-experts.
	Indeed, the use of logically based specifications is hard and error-prone~\cite{menghi:dwyer1999patterns}.

	Therefore this paper proposes \emph{strategy patterns}: an extension of the specification patterns for the CSP domain, recently proposed by Menghi et al.~\cite{menghi:Menghi:2018:PSP:3183440.3195044,menghi:PSALM}, that considers multi-agents CPS.

	\subsection*{Strategy Patterns}
	Specification patterns are a well-known software engineering solution for defining  requirements that are both human-comprehensible and machine-processable~\cite{menghi:dwyer1999patterns}.
	Patterns map recurrent human-comprehensible specification problems
	into machine-processable solutions.
	They include a usage intent, known uses, and relationships to other patterns, which facilitate
	humans in specifications, and
	a template  specification in a target logic.
	Extensions of the original patterns  of Dwyer et al.~\cite{menghi:dwyer1999patterns} have been proposed in the literature.
	An extension of the specification patterns tailored for the CSP domain has been recently proposed~\cite{menghi:Menghi:2018:PSP:3183440.3195044,menghi:PSALM}.
	However, it does not explicitly consider strategies, which, as discussed in the previous section, are  primary concerns in multi-agents CPS.

	Strategy patterns extend the property specification patterns for robotic missions~\cite{menghi:Menghi:2018:PSP:3183440.3195044,menghi:PSALM} by considering multi-agents CPS applications.
	This paper envisions two main  classes of strategy patterns that are tailored for cases in which agents are \emph{collaborating} and \emph{competing} for achieving their  goals.
	Within each of these classes, patterns capture recurrent mission specification problems.
	For example, the \emph{divide and conquer collaborative patrolling} pattern extends the patrolling pattern recently proposed by Menghi et. al.~\cite{menghi:Menghi:2018:PSP:3183440.3195044}.
	Given a set of locations to be patrolled by a set of agents, it assigns disjoint subsets of locations to the agents that must independently patrol them.
	Viceversa, in the \emph{unite and conquer collaborative patrolling} pattern, all the robots must patrol all the locations within  a given set.
	LTL is proposed as a target logic for the  template mission specifications.
	Indeed, LTL is strongly used in the CSP domain~\cite{menghi:luckcuck2018formal}, it has been used in the specification patterns for robotic missions, and it is also supported by several planners (e.g.,~\cite{menghi:FM2018}).

\end{conf-abstract}
\newpage
\begin{conf-abstract}[]
	{Challenges on the Way Towards Explainable Behavior of\\Adaptive Cooperative Mobile Systems}
	{Christoph Sommer}
	{Heinz Nixdorf Institute and Dept. of Computer Science\\ Paderborn University, Paderborn, Germany\\{sommer@ccs-labs.org}}
	\indexauthors{Sommer!Christoph}
	\noindent
	We live in the age of mobile computer-controlled systems.
	Car and truck manufacturers are working on autonomous vehicles, removing a central human element from the control loop that governs road traffic and allowing a computer to directly process and act upon sensor data.
	Such computer-controlled cars and trucks will form the fabric of ground transportation in future smart cities.
	Similarly, in the air, modern multirotor helicopter systems are moving from employing computers for no more than simple flight stability calculations to fully autonomous flight modes, targeting both goods delivery and passenger transport.
	
	The true evolutionary leap of such systems, however, will come not when the human operator is simply replaced by a computer (though this in itself will be a remarkable feat).
	Rather, it will come at the point where such computer-controlled mobile systems move from targeting autonomy to embracing cooperation, forming \emph{Cooperative Mobile Systems}.
	
	Cooperation of such mobile systems can take all forms from simple sharing of sensor data, to communicating intent, to cooperative planning, to sharing experiences, and to cooperative learning.
	Many of the challenges of cooperative mobile systems are rooted in classical disciplines of vehicular networks, from challenges in \emph{wireless networking, to system design, and to the simulation of such systems}:
	Cooperative mobile systems resemble a wireless network of extreme topology dynamics and scale, using heterogeneous communication technologies.
	
	
	\begin{figure}
		\centering
		\begin{tikzpicture}
		
		\node [circle, minimum size=2cm, text width=1.5cm, align=center, inner sep=0pt] (ex) {explainable cooperative mobile systems};
		
		\node [draw, circle, minimum size=5.75cm, text width=2cm, align=center, label=below:vehicular networking] (a) at (ex.south) {};
		\node [draw, circle, minimum size=5.75cm, text width=2cm, align=center, label=above:control theory] (b) at (ex.north) {};
		\node [draw, circle, minimum size=5.75cm, text width=2cm, align=center, label=right:software engineering] (c) at (ex.east) {};
		\node [draw, circle, minimum size=5.75cm, text width=2cm, align=center, label=left:machine learning] (d) at (ex.west) {};
		
		
		\end{tikzpicture}
		\caption{Explainable Cooperative Mobile Systems as an intersection of the fields of vehicular networking, control theory, machine learning, and software engineering.}
		\label{fig:intersection}
	\end{figure}
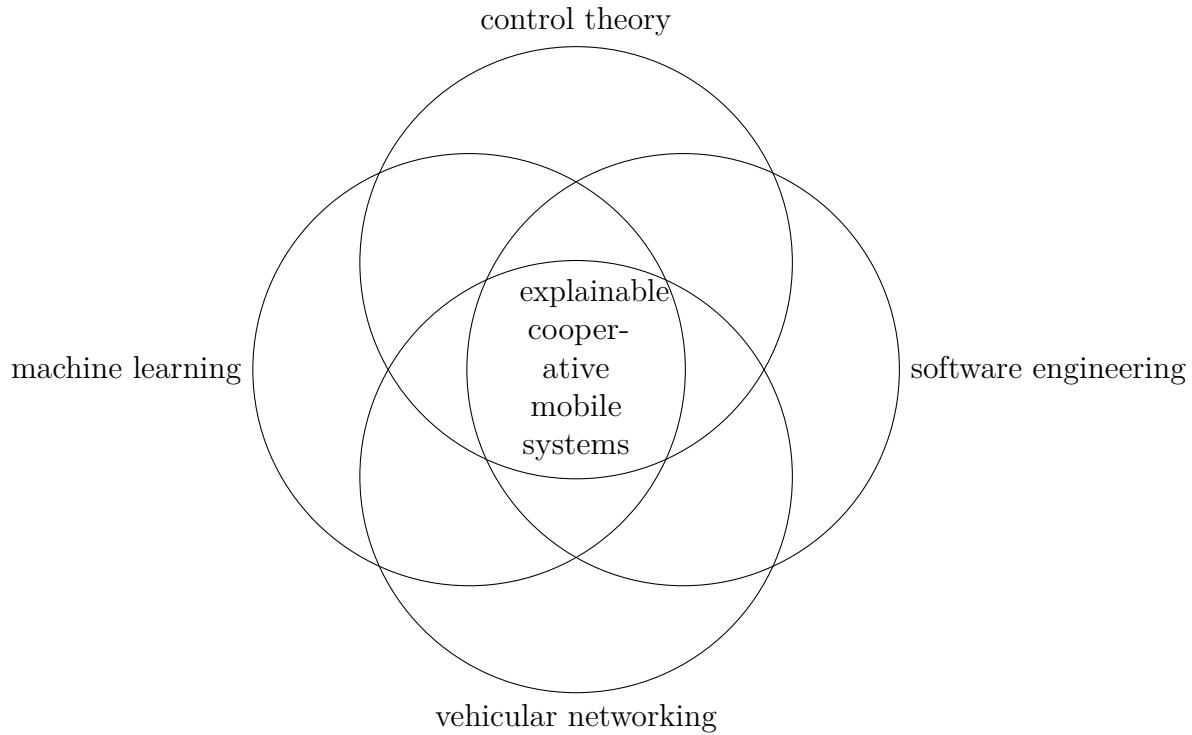

	If we look at applications like platooning (vehicles following at tight distances, closely coordinating their control input for longitudinal movement) or intersection control (vehicles coordinating for collision avoidance in a shared space without sacrificing on throughput), however, it becomes obvious that new challenges are emerging.
	
	First, we are now looking at a field of research where control theory meets wireless networking.
	Here, issues range from modeling uncertainty in control inputs to distinguishing simple wireless outages from misbehaving participants of the network, and from outside attacks.
	
	Second, autonomous control of vehicles is often tightly interwoven with machine learning approaches (both offline and online, to support self-adaptive behavior) thus adding another dimension to an already complex field.
	
	Finally, and taking both challenges together, because of the huge complexity of such systems, issues of \emph{explainability} will play a central role in their design and operation.
	Was a sudden brake maneuver triggered by a network outage?
	Was it the result of trust in the network participants having been eroded by unexpected behavior of one of the nodes?
	Was it the result of the driving situation being recognized as a potential hazard by a collaborative learning process?
	(Not) having a good answer to questions like this can easily be imagined to make-or-break real-world deployment of cooperative driving -- from the perspectives of all of potential users, of regulatory bodies, and of law enforcement.
	
	Tackling these challenges requires the pooling of resources from experts in all of the concerned fields (graphically illustrated in Figure~\ref{fig:intersection}): from wireless networking, simulation, control theory, machine learning, and software engineering.
	
\end{conf-abstract}
\newpage
\begin{conf-abstract}[]
	{Explainability of Provably Safe and Live Distributed\\ Controllers for Autonomous Car Manoeuvres}
	{Maike Schwammberger}
	{Department of Computing Science\\University of Oldenburg, Oldenburg, Germany\\schwammberger@informatik.uni-oldenburg.de}
	\indexauthors{Schwammberger!Maike}
	\noindent
	\subsection*{Motivation: What needs to be explained?}
	During the last years, driving assistance systems and fully autonomously driving cars are increasingly capturing the market worldwide. As autonomous car control involves both dynamic and discrete aspects, these systems are often implemented by Cyber-Physical Systems.

	It is of the utmost importance to ensure certain hard functional properties of autonomous cars, e.g. safety, meaning collision freedom with other traffic participants at any time. However, autonomous cars are complex, generally distributed and mobile, as well as more or less decentralized acting and possibly also communicating agents. Thus, analysing safety properties and explaining the behaviour of an autonomous car is a challenging task.

	\subsection*{Explainability of CPS: What is my expertise to the seminar topic?}
	One solution that uses logical reasoning about safety of traffic situations is the \emph{Multi-lane Spatial Logic (MLSL)} approach \cite{schwammberger:HLOR11}. While \cite{schwammberger:HLOR11} focuses on safety of car manoeuvres in freeway traffic, extensions, e.g. to urban traffic \cite{schwammberger:Sch17, schwammberger:Sch18-TCS}, have been proposed.

	In MLSL, we detach pure logical reasoning on spatial aspects of traffic situations from car dynamics \cite{schwammberger:ORWprocos17}. With this, it is possible to have concise logical statements about traffic situations with a clear formal semantics. The overall approach can be divided into three central parts:

	\begin{itemize}\itemsep0em
		\item \textbf{Abstraction:} An Abstract Model of real-world road structures, cars and their perception of other traffic participants. The abstract model includes a view, where only cars in an area around an active \emph{ego car} are considered for reasoning about safety of that car. This significantly simplifies the state space that needs to be considered and is motivated by reality as only cars around the ego car might endanger its safety.
		\item \textbf{Logic:} Reasoning about specific traffic situations in the Abstract Model, precisely in the view around the ego car. In MLSL, we formalise safety of ego by the formula
		\begin{align*}
		\text{\emph{Safe}}\,(\mathit{ego}) \;\equiv\; \neg \exists c \colon c \neq \mathit{ego} \land \langle re (\mathit{ego}) \land re (c) \rangle\text{,}
		\end{align*}
		where the atom $re (\mathit{ego})$ formalises the \emph{reservation} of the ego car, which is the space the car occupies on the road at one moment.
		\item \textbf{Car controllers:} Extended timed automata controllers use the logic to implement protocols for traffic manoeuvres that are provably live and safe. The safety formula holds invariantly in our traffic manoeuvre controllers.
	\end{itemize}

	\subsection*{Analysability: What is already explainable about MLSL?}
	The strength of our approach is its concise formal semantics of both logic and controllers. While a network of the several controllers running in parallel can be very big and difficult to understand manually, wanted or unwanted behaviour of our controllers can be analysed and explained during design-time, e.g. by using model checking approaches. For this, recent work was done in \cite{schwammberger:Sch18} on analysing safety and liveness aspects of the MLSL car controllers using the UPPAAL model checker \cite{schwammberger:BDL04}. We use \emph{Observer} automata to verify certain system behaviour, e.g. $\square\; \text{\emph{Safe}}\,(\mathit{ego})$ (``It is globally \emph{Safe} for \emph{ego}'') during simulated system runs. However, only a limited number of cars can be considered with UPPAAL and only specific queries are possible due to the UPPAAL query language.

	\subsection*{Outlook and seminar inputs: What is yet to be done?}
	As stated in the previous paragraph, our approach allows for analysability and thus explainability of certain system properties during design-time. However, our purely formal approach comes at the cost of a high level of abstraction and it is the question which results actually scale to reality. While the abstractions are neccessary and absolutely reasonable to enable pure logical reasoning, it needs to be considered which steps we can take to weaken some of them without losing expressiveness of our approach.

	Also, we currently only examine some selected important system properties for our MLSL controllers (safety, liveness, fairness). However, in the seminar we thoroughly discussed \emph{what} actually needs to be explained about a system and \emph{how} this could be managed. Moreover, in one breakout-group, we discussed a monitoring and analysis framework to enable self-explainability of certain system decisions and behaviour also \emph{during run-time} (see results of resp. breakout-group in Sect.~\ref{group:self-explainable-systems}). It would be quite interesting to apply these group results to our controllers. As we use extended timed automata for our controllers, their possible behaviour is encoded by their operational semantics (i.e. a transition system). One configuration in the semantics of the controller comprises a current location and valuation of all used variables. An example for this is the behaviour ``preparing a crossing manoeuvre'', which is active in the locations $q_1$, $q_2$ and $q_3$ of the crossing controller from \cite{schwammberger:Sch18-TCS}.

\end{conf-abstract}
\newpage

\begin{conf-abstract}[]
	{Explainable Quality Assurance of Behavioral Requirements}
	{Thomas Vogel}
	{Software Engineering Group\\Humboldt-Universität zu Berlin, Berlin, Germany\\thomas.vogel@informatik.hu-berlin.de}
	\indexauthors{Vogel!Thomas}
	\noindent
	\subsection*{Context: Safe.Spec}
	
	Imprecise, incorrect, and inconsistent specifications of requirements often cause faults in the software and result in increasing costs for developing and maintaining the software.
	In the \textit{Safe.Spec}\footnote{Safe.Spec is sponsored by the German Federal Ministry of Education and Research under Grant No. 01IS16027.} project, we develop an integrated tool for the quality assurance of behavioral requirements, that particularly:
	\begin{enumerate}\itemsep0.25em
		\item supports requirements engineers and domain experts in formally modeling behavioral requirements as scenarios expressed by UML sequence diagrams. For this purpose, we provide a catalog of behavior and interaction patterns (e.g., send and reply). A requirements engineer or domain expert instantiate and compose such patterns to create a scenario.
		\item automatically composes the modeled scenarios to an overall specification of the requirements that allows us to analyze the interplay between individual requirements. The overall specification is expressed by a network of timed automata for the UPPAAL\footnote{\url{http://www.uppaal.org/}} tool. To achieve the composition, we extend an existing algorithm~\cite{vogel:Uchitel:2003} by adapting it to the technical domains of UML and UPPAAL.
		\item supports requirements engineers and domain experts in formalizing properties to be verified on the overall specification of the behavioral requirements. Due to the difficulty and error-proneness of specifying properties in a temporal logic, we use the property specification patterns collected and proposed by Autili et al.~\cite{vogel:Autili:2015} that allows us to define properties in natural language using a structured English grammar. 
		\item verifies the defined properties against the overall specification of the requirements using UPPAAL. For this purpose, the Safe.Spec tool automatically translates the properties defined in natural language to Timed Computational Tree Logic (TCTL) and observer automata that serve together with the overall requirements specification as an input to UPPAAL. The translation connects the property specification patterns~\cite{vogel:Autili:2015} to UPPAAL taking UPPAAL's limited support of TCTL into account.
	\end{enumerate}
	
	Users of the Safe.Spec tool iterate over these four steps to specify and verify behavioral requirements. We anticipate that most faults in the specification will be avoided by formally modeling the requirements, in contrast to describing the requirements in natural language, while subtle faults will be detected by the verification. Thus, Safe.Spec contributes to the quality assurance of requirements and to the formalization of high-quality requirements.
	
	Focus of the Safe.Spec project are automotive systems that become more and more autonomous and open, and therefore require a high quality, for instance, with respect to safety, starting from the beginning of the development process with the requirements.

	\subsection*{Explainability in Safe.Spec}
	
	In the Safe.Spec project, we encounter at least three explainability problems:

	\begin{enumerate}\itemsep0	em
		\item The scenarios/requirements modeled with UML sequence diagrams should be textually described and explained. On the one hand, a textual description of the requirements might be needed due to legal reasons. On the other hand, it supports domain experts and novice users with little experience of modeling with UML in creating and understanding the scenarios.
		In the Safe.Spec project, we aim for automatically generating a textual description for each scenario that is derived from the behavior and interaction patterns used for composing the scenario.
		\item The verification properties expressed in TCTL and observer automata should be described and explained textually. For this purpose, the structured English grammar for creating the properties is used to generate a description of each property in natural language.
		\item The results from the verification performed on a network of timed automata (NTA) should be lifted and explained at the abstraction level of the scenarios (UML sequence diagrams). This is critical since users specify and change the requirements by means of scenarios while the NTA is automatically obtained by composing the scenarios. Thus, users interact with the scenarios and not directly with the NTA. Consequently, a counter example obtained for the NTA should be traced back to the UML sequence diagrams so that users can understand and relate it to the scenarios. This would provide actionable hints on how to change the scenarios to take the verification results into account. In this context, a research question is the selection of a suitable counter example. While UPPAAL provides either a \textit{shortest}, \textit{fastest}, or \textit{some} counter example, a notion of \textit{explainable} counter examples is missing.
	\end{enumerate}
	\vspace{-1em}
	
	To address these explainability problems, at least knowledge about model transformation, traceability, formal methods, and model checking is needed. To obtain reasonable and usable explanations, including domain knowledge seems to be a promising direction.

\end{conf-abstract}
\newpage

\begin{conf-abstract}[]
	{Towards Explainable RE Tools}
	{Andreas Vogelsang}
	{TU Berlin, Germany\\ {andreas.vogelsang@tu-berlin.de}}
	\indexauthors{Vogelsang!Andreas}
	\noindent
\subsection*{Introduction}
	Many Requirements Engineering tasks are nowadays supported by tools that either check the quality of manual RE work or perform RE tasks completely automatic. Examples are requirements categorization~\cite{vogelsang:winkler2016automatic}, prioritization~\cite{vogelsang:Perini.2013}, trace link recovery~\cite{vogelsang:Hayes03}, or detection of language weaknesses~\cite{vogelsang:Femmer16}. The increasing abilities of these tools is driven by the availability and accessibility of complex technologies. RE tools make use of advanced natural language processing techniques~\cite{vogelsang:Femmer16}, information retrieval mechanisms~\cite{vogelsang:Hayes03}, and machine learning (e.g., by  artificial neural nets~\cite{vogelsang:winkler2016automatic}).

	Despite the complex technologies used, RE tools are very appealing to practitioners because most of the technology is hidden from the user. However, when tools produce results that a user finds strange or that a user cannot explain, tools often fail to give evidence or hints \emph{why} it made this decision and what the consequences are. Moreover, for some of the complex technologies used it may even be impossible to provide reasons for some decisions. For example, it is very hard to explain why a neural net makes a specific decision.

	A special property of RE tools is that they are almost never used in a fully automated context. Most of the times, RE tools are part of processes, where they support a human analyst in performing tasks or reviewing work products. Therefore, we argue in this paper that more research is needed towards \emph{explainable} RE tools. An explainable RE tool is able to provide rationales or indication for the decisions that it makes.
	Moreover, we argue that RE tools should also provide \emph{actionable} results. A result is actionable if the tool can provide hints or recommendations on what could be done to improve or change a situation.

	We use the following informal definitions of these terms: An \emph{explainable} tool provides hints or indication on the rationale why the tool made a decision. An \emph{actionable} tool provides hints or indication on how the user can influence the decision by changing the processed data.
	In our experience, most tools and approaches reported in literature are not \emph{explainable} or not \emph{actionable}.

	\subsection*{Towards Explainable RE Tools}
	In the past, we made some efforts to make our RE tools explainable and actionable. Here, we provide an example.
	We have developed an automated approach to differentiate requirements from non-requirements (information) in requirements documents~\cite{vogelsang:winkler2016automatic}. At one of our industry partners, it is the document author's task to manually label all elements of a requirements document as either \emph{requirement} or \emph{information}. Our approach uses an artificial neural net that is trained on a large set of well-labeled requirements documents. After the training, the neural net is able to classify text fragments as one of the two classes. We use this approach to check the quality of this classification in existing documents. To make the decisions of the tool explainable, we have developed a mechanism that traces back the decision through the neural net and highlights fragments in the initial text that influenced the tool to make its decision~\cite{vogelsang:Winkler.2017}. As shown in Fig.~\ref{vogelsang:fig:tool}, it appears that the word ``must'' is a strong indicator for a requirement, whereas the word ``required'' is a strong indicator for an information. While the first is not very surprising, the latter could indicate that information elements often carry rationales (why something is \emph{required}).

	\begin{figure}
		\centering
		\includegraphics[width=\columnwidth]{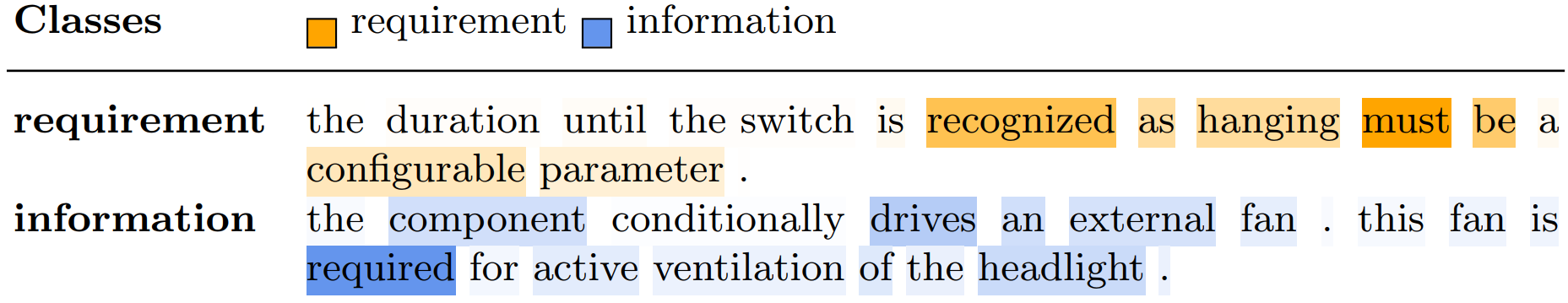}
		\caption{Explanation of the tool decision}
		\label{vogelsang:fig:tool}
	\end{figure}

	\subsection*{Beyond Explaining: Insights through Tools}

	While we see increased acceptance of RE tools as the main benefit of an explainable and actionable focus, we also envision that research towards explainable RE tools may also increase our understanding on how we write and understand requirements. A good example is the use of neural networks. While it is hard to comprehend why a neural net makes specific decisions, recent research has shown that it is not impossible to analyze the inner structure of neural nets to get deeper insights into the characteristics of the learned instances. For example, Bacciu~et~al.~\cite{vogelsang:Bacciu.fun} have used so called auto encoders to force a neural net to focus on the ``essential'' characteristics of jokes. By training the neural net behind the auto encoder on thousands of jokes given as texts, the inner structure of the neural net had to focus on those specifics in a text that are characteristic for jokes. By analyzing the inner structure, the authors were able to identify ``regions'' in which the net allocates specific text fragments that it classifies as the joke's punchline or ``dirty'' jokes. We think that it is an interesting area of research to apply unsupervised learning techniques to large sets of requirements to learn more about the characteristics of requirements by analyzing the inner structure of the resulting networks.

\end{conf-abstract}
\newpage

\begin{conf-abstract}[]
	{Towards Explainable Robotics}
	{Andreas Wortmann}
	{Software Engineering\\RWTH Aachen University, Aachen, Germany\\		{wortmann@se-rwth.de}}
	\indexauthors{Wortmann!Andreas}
	\noindent
\subsection*{Motivation}
Robotics is increasingly taking over the aspects of our lifes: automated
vehicles bring us to work~\cite{wortmann:CNN} where we cooperate with
industrial~\cite{wortmann:NatGeo} or service robots~\cite{wortmann:ABH+17} until we have
dinner in a robotic restaurant~\cite{wortmann:Forbes}. Nonetheless, robotics is one
of the most challenging fields of cyber-pyhsical systems' research.
The successfuly deployment of non-trivial robotics systems demands for the
inter-disciplinary collaboration of experts from a variety of different
domains, including, \eg mechanical and electrical engineering, navigation,
artificial intelligence, manipulation, human-robot interaction (HRI) design,
and systems engineering. All of these domains traditionally employ (mental)
models to describe or prescribe parts of their views on the systems under
development. For instance, in mechanical engineering, these models may be
CAD geometries or differential equations, in path planning, these may be the
manipulator capabilities, in artifical intelligence, these may be aritficial
neural networks, in HRI these may be user interface models, and in systems
engineering, these may be SysML~\cite{wortmann:FMS14} diagrams. In the
multi-disciplinary engineering of robotics systems, these models, which
reify different ways to think about systems, must be aligned and integrated
carefully to ultimately enable production of hardware and software of a
robotic system. To prevent the costly and error-prone manual integration of
models through, \eg best-practices or modeling guidelines, research on
software language engineering~\cite{wortmann:HRW18} investigates making the languages
behind the different experts' (mental) models explicit and
operationalizable.

Explain is a three-place verb which requires that an \textit{explainer}
explains \textit{something} to a \textit{receiver}. A research roadmap
to explainable robotics needs to consider all three places: For instance, the
system itself or an (a posteriori deployed) external explainer might explain
the systems' behavior or its intentions to paricipating domain experts or
system users. In some domains with strong regulation, reason codes have been
established to explicate and standardize domain-specific explanations (such
as the SEPA reason codes~\cite{wortmann:SEPA} for European banking). In less
regulated and standardized domains, a priori prescription of explanations is
not feasible. Instead, the explanations are specific to the individual
systems and their capabilitiles.

Therefore, one of the key challenges in explainable systems for CPS is to
(semi)automatically deriving explanations based on the systems'
multi-disciplinary development and run-time models. This demands for
\begin{itemize}\itemsep0em
	\item \emph{Modeling languages to describe explanation (parts) at design
		time and at run time}. Instead of creating novel general-explanation
	languages, deriving domain-specific explanation languages, similar to
	domain-specific transformation languages~\cite{wortmann:HRW18}, derived from the
	languages used by the respective domain experts might facilitate and
	foster their application.
	\item \emph{Integration of explanation models of different domains across
		various levels of abstraction and throughout the complete CPS lifecycle}.
	System behavior and intentions depend on the interaction of various
	software modules that operate at different levels of abstraction.
	Where run-time models describe behavior abstractly and delegate behavior
	implementation details to imperative code, the explainer nonetheless
	must bridge this gap to produce complete explanations suitable to
	different roles.
	\item \emph{Receiver-specific and time-specific tailoring of explanation
		granularity}. Explanations to different receivers and different times of a
	system's lifecycle demand for specific focus and granularity. Whereas a
	software engineer at design time might be interested in the stack trace of
	a TCP communication failures, an HRI expert investigating system behavior
	at run time might be interested in the fact that the remote communication
	partner is unavailable. Instead of generating individual explanations for
	all possible roles and lifecycle phases, the appropriate derivation of
	explanation views might help to better understand a system's behavior or
	intentions.
\end{itemize}



Goal: providing them with English explanations of robots' behavior
-  If the robot can generate an explanation of its behavior, it will increase users' trust
- important: narrow down the explanations to what humans are interested in (too much infomation)
- investigating \& modeling what types of information are important across
different contexts

\end{conf-abstract}

\chapter[Participants]{Participants}

\begin{table}[htbp]
	\centering
	{\footnotesize
		\begin{tabular}{p{55mm}p{65mm}p{55mm}}

		= Dimitri Bohlender \newline RWTH Aachen, DE &
		= Joel Greenyer \newline Leibniz Universität Hannover, DE &
		= Lars Luthmann \newline TU Darmstadt, DE \\ \ \\

		= Markus Borg \newline RISE SICS, SE &
		= Holger Hermanns \newline Saarland University, DE &
		= Claudio Menghi \newline University of Luxembourg, LU \\ \ \\		
		
		= Mathias Blumreiter \newline TU Hamburg, DE &
		= Nico Hochgeschwender \newline German Aerospace Center (DLR), DE &
		= Mohammad Reza Mousavi \newline University of Leicester, UK \\ \ \\

		= Francisco Javier Chiyah Garcia \newline Heriot-Watt University, UK &
		= Jesper Karlsson \newline KTH Royal Institute of Technology, SE &	
		= Christoph Sommer \newline Paderborn University, DE \\ \ \\	
		
		= Simos Gerasimou \newline University of York, UK &
		= Narges Khakpour \newline Linnaeus University, SE &
		= Maike Schwammberger \newline University of Oldenburg, DE \\ \ \\
		
		= Christopher Gerking \newline University of Paderborn, DE &
		= Verena Klös \newline TU Berlin, DE &
		= Thomas Vogel \newline HU Berlin, DE \\ \ \\
		
		= Ilias Gerostathopoulos \newline TU München, DE &
		= Maximilian Köhl \newline Saarland University, DE &
		= Andreas Vogelsang \newline TU Berlin, DE \\ \ \\
		
		= Holger Giese \newline HPI, University of Potsdam, DE &
		= Malte Lochau \newline TU Darmstadt, DE &		
		= Andreas Wortmann \newline RWTH Aachen, DE \\ \ \\

\end{tabular}
}\end{table}

\begin{center}
	\includegraphics[width=.82\linewidth]{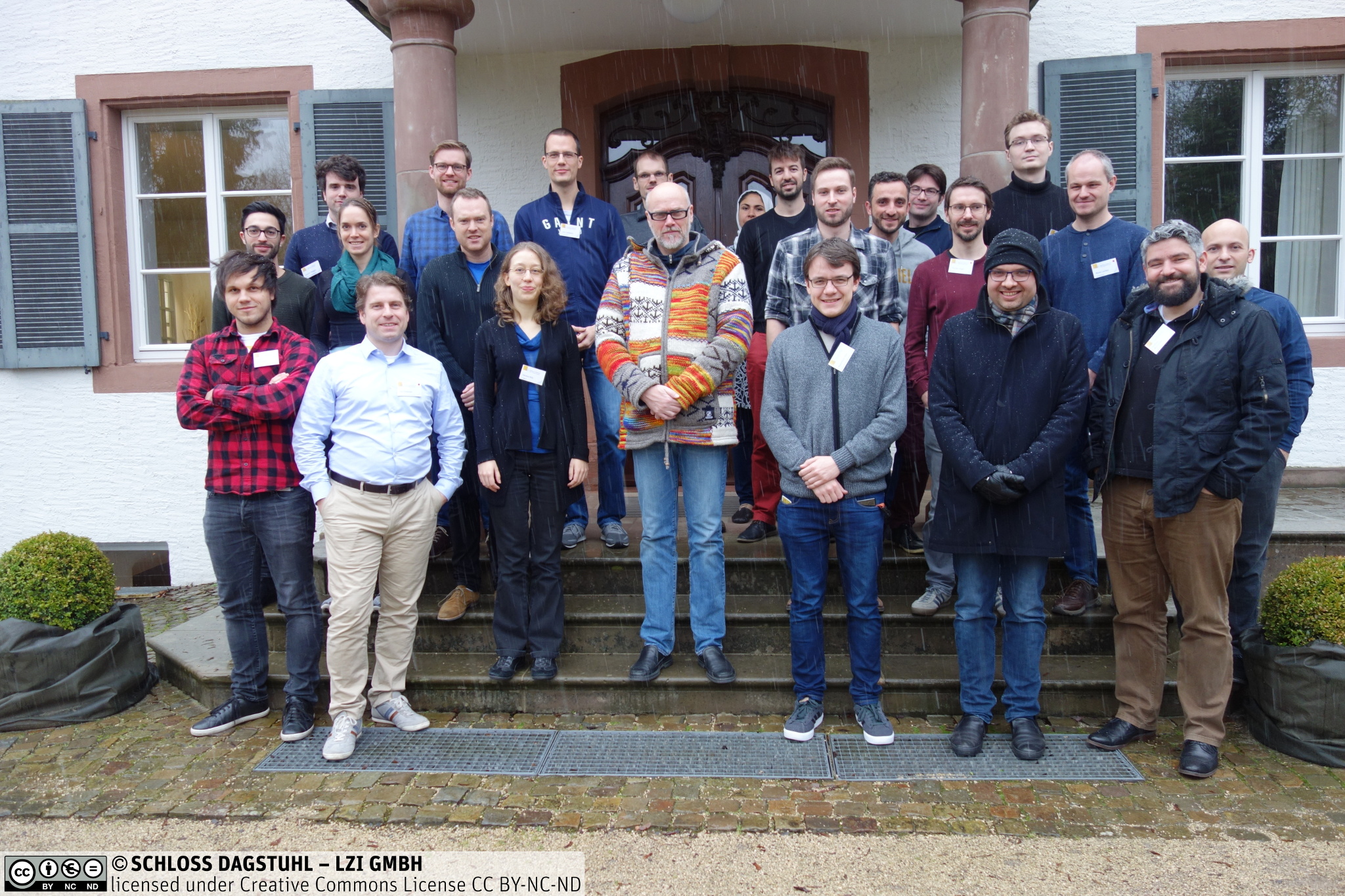}
\end{center}

\backmatter
\renewcommand{\indexname}{Author Index}
\printindex

\end{document}